\documentclass[trackchanges, twocolumn]{aastex7}
\usepackage{mathtools}

\defcitealias{2024ApJ...961..240K}{K24a}
\defcitealias{2024ApJ...961..241K}{K24b}
\defcitealias{2024ApJ...961..242R}{R24}
\begin{document}

\title{Minimum Energies and Magnetic Field Strengths of Edge-Brightened Compact Symmetric Objects}

\correspondingauthor{Tirth Surti}
\email{tsurti@caltech.edu}

\author[orcid=0000-0002-6369-6266]{Tirth D. Surti}
\affiliation{Cahill Center for Astronomy and Astrophysics, California Institute of Technology, 1216 E California Blvd, Pasadena, CA 91125,
USA}
\email{tsurti@caltech.edu}

\author[orcid=0000-0002-0610-2644]{Martijn S.\,S.\,L. Oei} 
\affiliation{Cahill Center for Astronomy and Astrophysics, California Institute of Technology, 1216 E California Blvd, Pasadena, CA 91125,
USA}
\affiliation{Leiden Observatory, Leiden University, PO Box 9513, 2300 RA Leiden, The Netherlands}
%\affiliation{Leiden Observatory, Leiden University, Niels Bohrweg 2, 2300 RA Leiden, The Netherlands}
\email{oei@caltech.edu}

\author[orcid=0000-0001-9152-961X]{Anthony C.\,S. Readhead}
\affiliation{Cahill Center for Astronomy and Astrophysics, California Institute of Technology, 1216 E California Blvd, Pasadena, CA 91125,
USA}
\email{acr@astro.caltech.edu}

\author[orcid=0000-0002-9545-7286]{Andrew G. Sullivan}
\affiliation{Kavli Institute for Particle Astrophysics and Cosmology, Department of Physics, Stanford University, Stanford, CA 94305, USA}
\email{ags2198@stanford.edu}
\begin{abstract}
Compact symmetric objects (CSOs) are subkiloparsec radio sources with two-sided emission about a core resulting from jets that are not relativistically beamed towards the observer. This relative simplicity makes them attractive targets to study the launching and evolution of relativistic jets. We use radio surveys and spatially resolved VLBA observations to estimate the minimum energies and magnetic field strengths of a subset of edge-brightened CSOs (CSO-2s). These are necessary to test models of CSO-2 formation via stellar capture and evolution via synchrotron cooling. By treating the observed X-ray emission of CSO-2s as inverse Compton emission from synchrotron and external photon fields, we estimate a mean departure from the minimum energy magnetic field strengths of ${\sim}2\times$, suggesting that CSO-2 lobes are close to minimum energy. Typical lobal minimum energy magnetic field strengths of $20$ mG suggest that once the jets shut off, luminous CSO-2s should fade at GHz frequencies within ${\sim}10^{3}$ years. We find that CSO-2 minimum energies are systematically larger than previously estimated. If luminous CSO-2s result from tidal disruption events, a majority would require the capture of massive stars $>1 \ M_{\odot}$ assuming jet launching efficiencies less than $100\%$.
%<1$.

%Edge-brightened CSOs (CSO-2s) have recently been argued to represent a distinct class of short-lived jet systems with a sharp size cutoff at ${\sim}500$ pc. The energies and rates of CSO-2s have also been used to support a tidal disruption event (TDE) formation scenario. We use radio surveys and spatially resolved VLBA observations to estimate the required minimum energies to test the TDE scenario and the synchrotron cooling lifetimes to constrain existing models of CSO-2 evolution.
\end{abstract}

\keywords{\uat{High-luminosity active galactic nuclei
}{2034} --- \uat{Extragalactic radio sources}{508} --- \uat{Extragalactic magnetic fields}{507} --- \uat{Compact objects}{288}}


\section{Introduction} 
Compact symmetric objects (CSOs)  are a class of compact radio-emitting active galactic nuclei (AGN), similar in morphology to the traditional Fanaroff-Riley type I and type II objects \citep{1974MNRAS.167P..31F}, but $10$--$10^{4}\times$ smaller and shorter lived. CSOs have long been studied using very long baseline interferometry (VLBI), providing the necessary resolution to resolve their symmetric radio lobes. Through an extensive study carried out by \citet{2024ApJ...961..240K, 2024ApJ...961..241K} (hereafter \citetalias{2024ApJ...961..240K} and \citetalias{2024ApJ...961..241K}), their definition has recently been redefined. In particular, the authors have defined CSOs to (i) exhibit two-sided radio structure symmetrically distributed about the center of activity (which need not be visible), (ii) have a largest projected length $<1$ kpc, (iii) not have strongly beamed emission towards the observer, and (iv) not have apparent superluminal motion $\beta_{\mathrm{app}} > 2.5$. Criterion (iii) is often concretized by the source exhibiting a lack of variability ($<20\%\ \mathrm{yr}^{-1}$) in conjunction with criterion (iv), which would suggest order unity lobal Doppler factors. An extensive literature search by \citetalias{2024ApJ...961..240K} resulted in the identification of 79 observed bonafide CSOs, and work done by \cite{Sheldahl_2025} since has identified an additional 65 CSOs meeting the classification criteria. Defining the spectral index $\alpha$ through $F_{\nu} \propto \nu^{\alpha}$, \citet{2024ApJ...977..195D} have shown that CSOs exhibit steep ($\alpha < -0.5)$ and flat ($\alpha \geq -0.5$) spectra between 5--8 GHz; CSOs are typically GHz-peaked sources, but as surveyed by \citet{1998PASP..110..493O}, not all GHz-peaked sources belong in the CSO class.

\cite{2016MNRAS.459..820T} showed that, like Fanaroff-Riley sources, CSOs typically fall into an edge-dimmed class (designated as CSO-1s) and an edge-brightened class (designated as CSO-2s). Using flux-limited complete samples of radio sources, \citetalias{2024ApJ...961..241K} argue that CSO-2s may constitute a distinct class of AGN, as suggested by a tentative sharp cutoff in the size distribution at around $500\,\mathrm{pc}$ and distinct redshift distribution compared to the jetted-AGN population. The former, in particular, suggests that CSO-2s are short-lived sources at surface brightness levels accessible by the Very Long Baseline Array (VLBA). CSO-1s were not analyzed due to their small occurrence rate in the complete samples. CSO-2s are further divided up into three classes (CSO-2.0--2.2), where \cite{2024ApJ...961..242R} argue that CSO-2s evolve from 2.0s (younger, higher luminosity) to 2.1s and 2.2s (older, lower luminosities), as modeled by \citet{10.1093/mnras/stae322} in a $(P, D)$, or luminosity--size, diagram. However, these simulations suggest a pile up of CSO-2s at large sizes and luminosities, which are not consistent with VLBA observations of CSOs (see Fig. 1 in \citet{2024ApJ...961..242R}, hereafter \citetalias{2024ApJ...961..242R}).

Consistencies with CSO-2 birth rates, lifetimes, and energies led \citetalias{2024ApJ...961..242R} to suggest that CSO-2s could be the result of tidal disruption events (TDEs) of a post--main sequence star of a few solar masses, as had first been suggested by \citet{1994cers.conf...17R}. Scaling from known CSO equipartition energies derived previously, \citetalias{2024ApJ...961..242R} estimated a range of equipartition energies from $10^{-4}\,\mathrm{M_{\odot}c^2}$ to $7\,\mathrm{M_{\odot}c^2}$ among the CSO-2 population. \citetalias{2024ApJ...961..242R} suggest that the low-energy CSO-2s can be triggered by the stellar capture of evolved ${\sim 1}\,\mathrm{M_{\odot}}$ stars, whereas the high-energy CSO-2s can be triggered by more massive stars, despite being rarer.
%consistent with the scenario of stellar capture of less massive stars, assuming typical black hole extraction efficiencies of $<1$.

In this paper, we conduct a more rigorous synchrotron minimum energy analysis of the CSO-2s identified in \citetalias{2024ApJ...961..240K}, \citetalias{2024ApJ...961..241K}, and \citetalias{2024ApJ...961..242R} to derive the characteristic component magnetic field strengths and minimum energies to constrain proposed models of their formation and evolution. We present two different ways of estimating these quantities, the first using individual components of CSO-2s and the second using the full component-sum radio spectrum. We then model the broadband AGN spectrum of coreless CSO-2s to evaluate whether the synchrotron and external photons from the AGN and host galaxy can well-explain their observed X-ray emission without deviating significantly from minimum energy.

%We also model the AGN spectrum of coreless CSO-2s, finding that inverse Compton emission of synchrotron and external photons from the AGN/host galaxy can well-explain their observed X-ray emission without deviating too far from minimum energy. With characteristic magnetic field strengths of ${\sim}20$ mG, CSO-2s should fade at VLBA frequencies on timescales comparable to their dynamical timescales once their jets turn off. We also find that CSO-2s exhibit systematically larger minimum energies than that suggested by \citetalias{2024ApJ...961..242R}, a majority of which are in excess of $1\,\mathrm{M_{\odot}}c^2.$ If these CSOs are several factors off from minimum energy, then the energy requirements to be produced by a TDE are even higher than previously suggested.

Throughout this paper, we assume a flat $\Lambda$CDM cosmology with $h=67.66$ and $\Omega_\mathrm{M,0} = 0.3111$ \citep{Planck12020}. In addition, we define the spectral index $\alpha$ so that $F_\nu \propto \nu^\alpha$.

\section{Data} \label{sec:data}

In this paper, we focus on conducting a synchrotron and inverse Compton emission analysis of CSO-2s to derive magnetic field strengths, minimum energies, and departures from thereof. This requires careful modeling the volume of the emitting regions of CSO-2s and their optically-thin synchrotron spectrum to obtain their bolometric synchrotron luminosities. The data collected to perform this analysis are motivated by two different methods we considered for estimating the CSO-2 lobal minimum energy conditions, each with their own merits and drawbacks. Because the data quality does not allow us to reliably distinguish them, we call any resolved island of flux (jet, knot, lobe, core, or hotspot) a component in this paper.
\begin{enumerate}
\item We use spatially resolved VLBA maps at two different frequencies in the optically thin part of the synchrotron spectrum and derive minimum energies and magnetic field strengths for each identifiable component that is not a core using the volumes and luminosities of each component.
\item We use lower-resolution radio surveys that generally do not allow for a separation of components, and use them to fit a broadband radio spectrum to estimate the component--sum synchrotron luminosity of the source. We use this spectrum, along with the sum of the VLBA-derived component volumes, to find the total minimum energy and averaged magnetic field strength for the source as a whole.
\end{enumerate}
While method 1 enables us to infer minimum energies and magnetic field strengths on a per-component basis, it can only be applied to CSO-2s that have two high-quality VLBI maps available for which components can be matched by eye. Furthermore, due to missing u-v coverage at the shortest baselines, VLBA maps may underestimate the flux density of components and produce steeper spectral indices when only taking the difference between two maps. Finally, without a full spectral decomposition, it is difficult to know whether the two frequencies are optically thin or partially absorbed for a given component. These concerns justify method 2, for which fitting to a full spectrum can better estimate the overall synchrotron luminosity of the source at the expense of losing the distinction of individual components. However, as discussed later in Section \ref{sec:min_e_res}, method 2 can suppress the energy contribution from large--volume low--surface brightness components that do not contribute significantly to the synchrotron luminosity.

For method 1, we collected calibrated images from the Radio Fundamental Catalog \citep[RFC;][]{Petrov12025}. Out of the \citetalias{2024ApJ...961..242R} sample of 43 sources with spectroscopic redshifts, we identified only 25 to have VLBA maps in the RFC showing unambiguous symmetric emission at two different VLBA frequencies that could be matched by eye. An additional 9 sources were identified to have symmetric components at only one frequency and were thus used for component volume estimation in method 2.
%in a single-frequency map
The remaining 9 sources belong in one of the following categories and thus were not considered in the following analyses: (i) they have no available RFC data; (ii) at least two symmetric components cannot be detected in any available RFC map; or (iii) elliptical gaussians are not a good approximation to the component morphology. Case (iii) occurs when there is a significant amount of diffuse or resolved-out emission. Altogether, we analyze 34 CSO-2s of which 16 are 2.0s, 10 are 2.1s, and 8 are 2.2s based on the \citetalias{2024ApJ...961..242R} classification.

To select VLBA maps, we first prioritize using those that are of good quality and within which symmetric components can be identified by eye. We then prioritize using maps that are taken as close as possible in time to more easily identify cores that dominate intrinsic source variability. For most sources, we use the frequently available S-band (2.3 GHz) as our lower-frequency VLBA map and C (5.0 GHz) or X-band (8.4 GHz) as our higher-frequency VLBA map. For sources that peak at higher frequencies or do not have resolved components at lower frequencies, we opt to use the U-band (15 GHz) MOJAVE observations \citep{Lister2018} as the higher-frequency map. Seven sources have maps that are separated by at least 2 years, but the inclusion or exclusion of these sources does not change our conclusions, and spectral indices of components are generally consistent with what is found in literature. In Table \ref{tab:rfc_meta} of Appendix \ref{sec:rfcVLBAmaps}, we list the frequencies and observation dates of the selected RFC maps per source.

For method 2, with the advent of all-sky radio surveys, we are able to measure the shape of the synchrotron spectra. We constructed radio spectra by collecting flux density measurements from literature catalogues of which the central frequencies span the $54$--$442,500\ \mathrm{MHz}$ range. We note that the catalogues span multiple epochs.

%While the catalogues span different epochs, we do not expect strong variations in spectral shape due to the low variability of CSOs.
%derive from source extraction performed on --> correspond to
Sorted by central frequency from low to high, these catalogues derive from the LOFAR LBA Sky Survey \citep[LoLSS;][]{deGasperin12023}, the VLA Low-Frequency Sky Survey redux \citep[VLSSr;][]{Lane12012}, the LOFAR Two-Metre Sky Survey \citep[LoTSS; e.g.][]{Shimwell12017, Shimwell12022}, the TIFR GMRT Sky Survey Alternative Data Release \citep[TGSS ADR;][]{Intema12017}, the Galactic and Extragalactic All-Sky MWA Survey \citep[GLEAM;][]{HurleyWalker12017}, the Westerbork Northern Sky Survey \citep[WENSS;][]{Rengelink11997}, the VLASS Commensal Sky Survey \citep[VCSS;][]{Polisensky12016, Clarke12016}, the Rapid ASKAP Continuum Survey Low \citep[RACS-low;][]{Hale12021}, the Rapid ASKAP Continuum Survey Mid \citep[RACS-mid;][]{Duchesne12024}, the NRAO VLA Sky Survey \citep[NVSS;][]{Condon11998}, the Faint Images of the Radio Sky at Twenty Centimeters \citep[FIRST;][]{Becker11995}, the Rapid ASKAP Continuum Survey High \citep[RACS-high;][]{Duchesne12025}, the VLA Sky Survey \citep[VLASS;][]{Gordon12021}, the Combined Radio All-Sky Targeted Eight GHz Survey \citep[CRATES;][]{2007ApJS..171...61H}, the VLA Calibrator Catalogue (VLA CC){\footnote{\url{https://science.nrao.edu/facilities/vla/observing/callist}}}, the \textit{Planck} Multi-frequency Catalogue of Non-thermal Sources \citep[PCNT;][]{Planck12018}, and the
ALMA Calibrator Catalogue \cite[ACC;][]{Bonato12019}.

For each survey, we list the central frequency and the point spread function (PSF) full width at half maximum in Table \ref{tab:surveys}. Uncertainties on the VLA calibrator flux densities are assumed to be 10\% for L--X bands, 20\% for U--K bands, and conservatively 30\% for Q following the broad guidelines on the calibrator site. If there is a quality flag `?' at any configuration or the quality flag is `X' across all configurations, the measurement is ignored. We assume a nominal 10\% uncertainty on the CRATES flux densities. We do not account for angular resolution differences across instruments. CSOs are generally compact ($\lesssim$100 mas in size) and the chosen surveys are either single--dish or arrays with short u--v spacings. Therefore, CSOs will generally be unresolved and if not, any extended diffuse emission would be accounted for. While ALMA can achieve resolutions where CSOs become resolved in the most extended configuration, image--plane model--fitting was done to model the flux of potentially resolved sources \citep{10.1093/mnras/sty1173}. Similarly, for the VLA CC, the quality cuts ensure that flux measurements were selected for sources at frequencies where the structure is known and can be used at minimum as a suitable phase calibrator. Additionally, by nature of the high VLBA surface brightness--limits (${\sim}1\,\mathrm{mJy/beam}$ at GHz frequencies), the selected CSOs are among the brightest at each frequency and are thus less likely to be affected by source confusion.

\subsection{Notes on Individual Sources}
We could not find 5 and 8 GHz measurements for several sources in the aforementioned surveys. Our searches through the literature provided 4.85 and 8.4 GHz flux densities for J0855+5751 and J1414+4554 and 4.85 GHz flux densities for J1159+5820, J1440+6108, J1816+3457, and J1434+4236 \citep{1991ApJS...75.1011G, 1992MNRAS.254..655P}; we assume uncertainties of 10\% if not given. For J1511+0518, we have added additional measurements up to 343 GHz from the ACC taken in 2021 following the publication of \cite{Bonato12019} as initial fits in the optically thin regime were not well--constrained. Finally, for J2022+6136, we require data from the Second Planck Catalogue of Compact Sources \citep{2016A&A...594A..26P} at 70 and 100 GHz to constrain the optically-thin part of the spectrum.

\subsection{High Energy Observations of CSO-2s}
A $1^{\prime}$ search around each selected CSO-2 in the Fermi LAT 14-year point source catalog \citep[\textit{4FGL-DR4}; ][]{2020ApJS..247...33A} shows no sources with detected gamma-ray emission. As discussed later in Section \ref{sec:inv_comp}, many CSO-2s exhibit detectable X-ray emission with \textit{XMM-Newton}, \textit{NuSTAR}, or \textit{Chandra} \citep[e.g., see][]{Gan_2025, 10.1093/mnras/stae2817} consistent with a single power law from 2--10 keV, with J0713+4349, J1407+2827, J1511+0518, and J2022+6136 additionally showing evidence of a broad Fe K$\alpha$ line in the 2--10 keV X-ray spectrum. Such sources may have more core activity and variability, and we consider the resulting minimum energies and magnetic field strengths by also excluding these sources in Section \ref{sec:min_e_res}. These sources are ignored in the inverse Compton analysis of Section \ref{sec:inv_comp} because the iron line suggests a non-negligible X-ray emission contribution from the core rather than the extended lobal components.

\section{Synchrotron Analysis}
\label{sec:anal}

For both methods, we apply the same formalism to estimate the synchrotron minimum energies and magnetic fields. If limits on the synchrotron spectrum are to be set by the minimum and maximum energies of the underlying particle energy distribution assumed to be a power law, then the characteristic synchrotron frequency (in SI),
\begin{equation}
\nu_\mathrm{c} = \frac{3eB}{16m_e}\gamma^2,
\label{eq:char_freq}
\end{equation}
which has been averaged over isotropic pitch angles, will set the lower and upper frequency cutoffs. In this paper, we assume a fiducial $\gamma_{\min}=10$ and $\gamma_{\max}=10^{5}$ for first-order Fermi acceleration \citep[as done for FR IIs in][]{10.1111/j.1365-8711.2000.03883.x}, so for typical CSO magnetic field strengths on the order of $10\,\mathrm{mG},$ the synchrotron spectrum will have cutoffs at $\nu_{\min}\simeq3\,\mathrm{MHz}$ and $\nu_{\max}\simeq300\,\mathrm{THz}.$ The bounds on the synchrotron spectrum are set by the magnetic field strength of the source and $\gamma_{\min}$ and $\gamma_{\max}$. We assume that the CSO-2s are continuously refreshing the electron population in their components. Even if the high-energy electrons cool faster than they can get replenished leading to a cooling break, their contribution to the overall energetics will be minimal. In Section \ref{sec:min_e_res}, we also discuss the effect of the electron $\gamma$ range on the minimum energy quantities, and a reduction of $\gamma_{\max}$ can act as fixing a cooling cutoff. Given the different formalisms often used to calculate the minimum energy conditions for CSOs, we clarify our procedure in Appendix \ref{sec:deriv}. Equations \ref{eq:b_min} and \ref{eq:e_min} give the explicit formulae for the magnetic field strength and minimum energy, respectively. The main observables required to use these equations are (i) the volume of the emitting region and (ii) the synchrotron luminosity at a reference optically--thin frequency, which are explained in Sections \ref{sec:ind} and \ref{sec:spec_anal} for each method. Similar derivations of synchrotron minimum energies and magnetic field strengths can be found in \cite{1956ApJ...124..416B, 2005AN....326..414B, 2011hea..book.....L}.

Our methodology primarily differs from \citetalias{2024ApJ...961..242R} in that we explicitly model the volumes of the emitting regions from the VLBA maps, and use the intrinsic bolometric synchrotron luminosity determined by integrating an extrapolation of the optically thin part of the spectrum (between the low- and high-frequency synchrotron cutoffs) as a proxy for the total particle content.
%(rather than the observed peak radio luminosity) 
In contrast, \citetalias{2024ApJ...961..242R} use the scaling relation $U_\mathrm{eq} \propto L_{\nu,\mathrm{peak}}^{4/7}D_{\mathrm{lobe}}^{4/7}$ to estimate the maximum CSO-2 equipartition energy, assuming a peak observed radio luminosity $L_{\nu,\mathrm{peak}} \simeq 3\times10^{27}\,\mathrm{W/Hz}$ and mean lobe size $D_{\mathrm{lobe}} \simeq 50\,\mathrm{pc}$.
We argue that using the intrinsic synchrotron luminosity best represents the total particle content within the CSO components; the peak radio luminosity, instead, is sensitive to synchrotron self-absorption, and thereby discounts particles in the (frequency-dependent) optically thick interiors of CSO components.
%Furthermore, we set the bounds on the intrinsic synchrotron spectrum equal to the 
Furthermore, our spectral integration domain is the characteristic synchrotron frequency range of electrons with Lorentz factors between some fixed minimum $\gamma_\mathrm{min}$ and maximum $\gamma_\mathrm{max}$, rather than the frequency range presently accessible to radio surveys (i.e. 54 MHz--400 GHz).
%because synchrotron self-absorption takes away radiative energy through re-absorption by particles that is thus not accounted for when using just the peak radio luminosity.}

We note that the proton (or more generally, hadronic) contribution to the particle energy density $K$ and the volume filling factor $\eta$ remain unconstrained and are assumed to be zero and one respectively. We also note that the assumption $K=0$ is consistent with that of \citetalias{2024ApJ...961..242R}. Additionally, VLBA maps, especially those at higher frequencies, may underestimate volumes $V$ due to lower sensitivity to extended emission and lack of short-spacing u-v coverage. In a minimum energy analysis, uncertainties in $K$, $\eta$, and $V$ are suppressed in the magnetic field calculation given that $B_\mathrm{ME} \propto \left(\frac{1+K}{\eta V}\right)^{1/(3-\alpha)}$, where typically $\alpha < -0.5$. Therefore, order of magnitude uncertainties in $(1+K)/\eta V$ will correspond to uncertainties of order unity on $B_{\min}$. However, the minimum energy $E_{\min}\propto \left(1+K\right)^{2/(3-\alpha)}(\eta V)^{(1-\alpha)/(3-\alpha)}$ is dependent on larger powers of the unconstrained variables and thus more affected by systematic assumptions on them. For example, an order of magnitude uncertainty on $\eta V$ alone can produce uncertainties of a factor of ${\sim}3$ on $E_{\min}.$

%However, volume filling factors in regions of diffuse emission are likely $\ll1$ and higher frequency VLBA observations may probe emission regions that are more certain to have $\eta\approx 1$.

\subsection{Individual Component Analysis}\label{sec:ind}

\begin{figure*}[htp!]
    \centering
    \includegraphics[width=\linewidth]{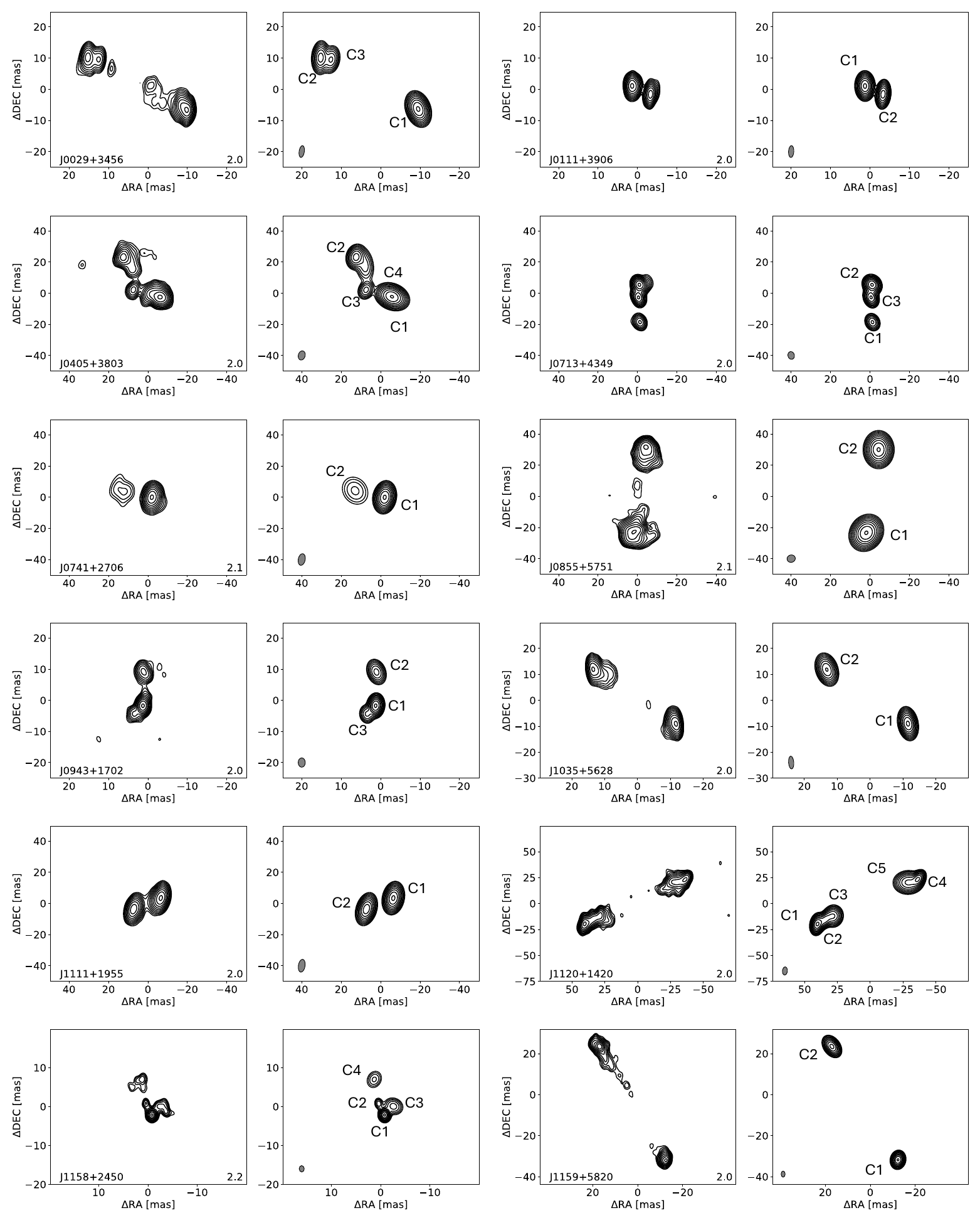}
    \caption{Low-frequency VLBA maps \textit{(left)}, directly compared with component fits of CSO-2s in the sample \textit{(right)}.
    %Component fits of CSO-2s in the sample, directly compared to the low-frequency VLBA map.
    For each source, we use 10 contour levels sampled in log space from 1\% to 90\% the maximum, with the exception of J1120+1420, J1158+2450, J1159+5820, and J1244+4048 having a lower limit set to 2.5\% and J1440+6108 set to 4\% to reduce map noise and J0741+2706 set 0.5\% to bring out the faint eastern component. Fitted components have been labeled and fits of fluxes and spectral indices are given in Table \ref{tab:ind_results}. We have labeled the CSO-2 sub-classifications in the bottom right of each VLBA map panel.}
    %containing the reconstruction.}
    \label{fig:component_fit}
\end{figure*}

\begin{figure*}
    \centering
    \includegraphics[width=\linewidth]{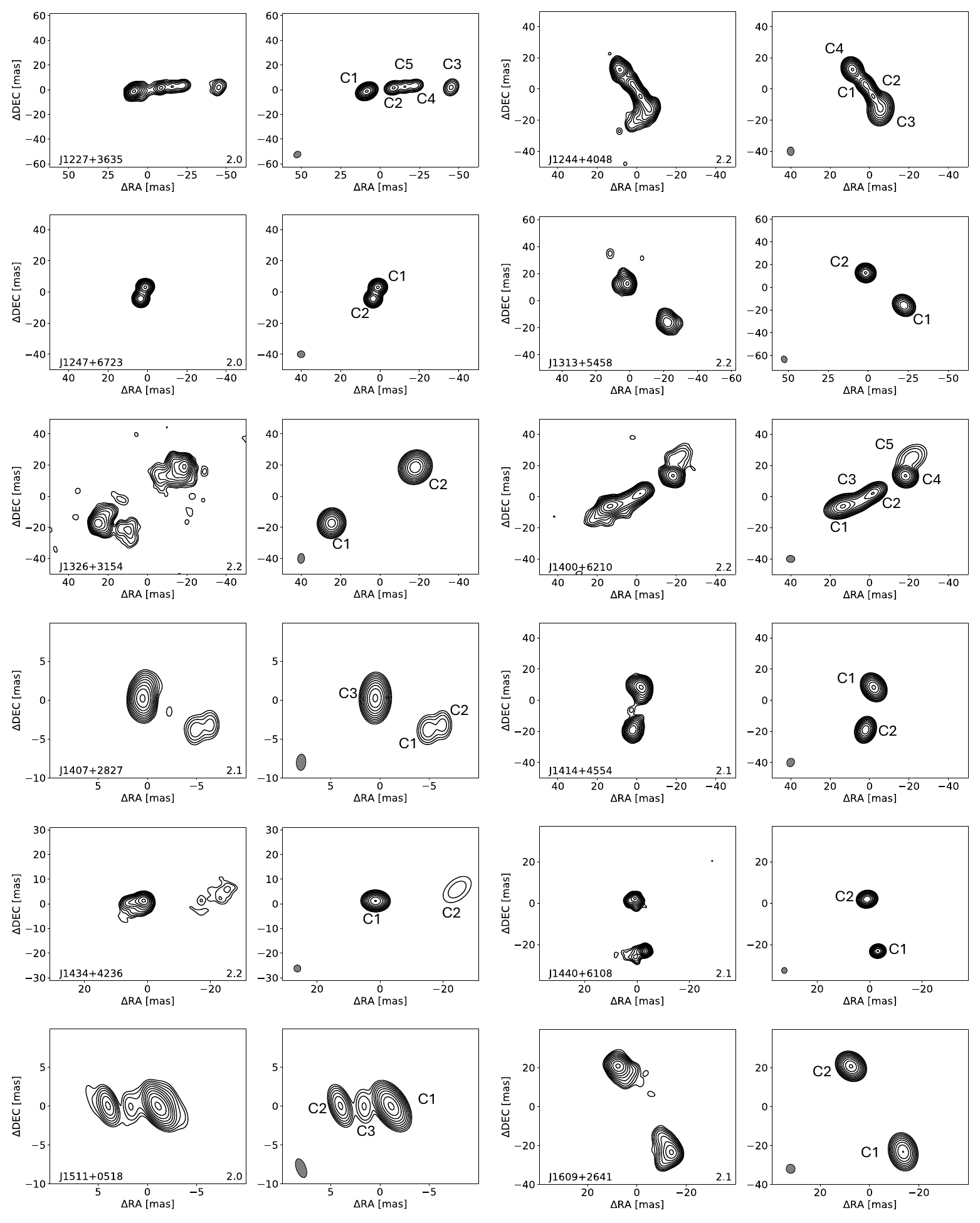}
    \centerline{\small (Continued.)}
\end{figure*}

\begin{figure*}
    \centering
    \includegraphics[width=\linewidth]{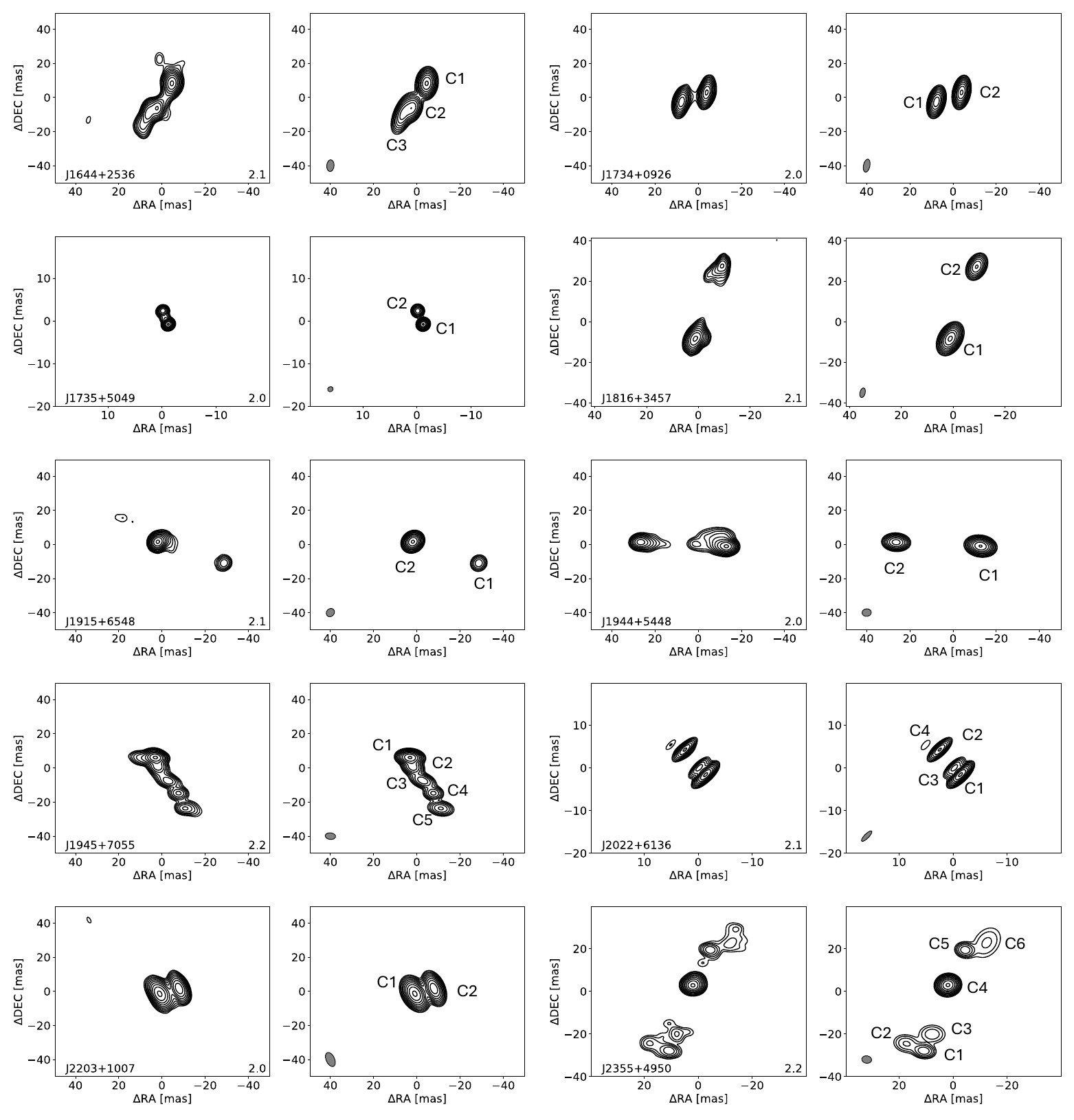}
\end{figure*}

For each of the 25 sources identified to have high quality VLBA maps at two frequencies, we use the radio source finding tool \textsc{Aegean-Tools} \citep{2012MNRAS.422.1812H, 2018PASA...35...11H} to identify all component (and core) locations. We use a nominal ``seeding'' index of $10\sigma$ and ``flooding'' index of $4\sigma$ above the map RMS (which determines the boundary of the ellipse); consequently, all the components considered in this analysis have a signal-to-noise ratio (SNR) ${>}10$. Background RMS values are estimated using the program \textsc{BANE} from \textsc{Aegean-Tools}. We match components across both the VLBA maps for spectral index computation. To improve component matching, we convolve the higher frequency map to the resolution of the lower frequency map (if possible and if components do not get blurred together).

Because \textsc{Aegean-Tools} does not reliably return uncertainties on all fits, we use a Markov Chain Monte Carlo (MCMC) approach via the \textsc{emcee} package \citep{2013ascl.soft03002F} to obtain uncertainties on the component volumes and flux densities. For each component or group of components that overlap in the image plane identified by \textsc{Aegean-Tools}, we use the output of \textsc{Aegean-Tools} as a starting point to fit elliptical Gaussians with freely varying major and minor FWHMs, position angles, and amplitudes; the positions remain fixed. The MCMC samples the intrinsic major and minor FWHMs first and then convolves them by the beam to match the VLBA map, so that uncertainties on the true component size can be obtained. We use a Gaussian likelihood with the derived background RMS being the per-pixel intensity standard deviation. We use these samples to propagate uncertainty into the flux density, spectral index, and volume calculations. Due to the higher sensitivity to extended flux, we use the intrinsic Gaussian fit to the lower-frequency VLBA maps to estimate the component volume. We assume that the components are ellipsoids, that the intrinsic projected major and minor FWHMs equal the intrinsic semi-major and semi-minor axes, and that the third axis is the average of the two projected axes. For all sources, we fit to the same maps used for \textsc{Aegean-Tools}.

Figure \ref{fig:component_fit} shows maps at the low frequency of all CSO-2s in the sample (including those with single VLBA maps), with the fitted components labeled. For morphologically simple sources with two elliptical components like J1734+0926, the Gaussians can well explain the emission. However, for more morphologically complex sources like J1159+5820, where there is faint diffuse emission, and J0029+3456, whose central core was not detected in the higher frequency X-band map and thus not covered by the model, the full emission decomposition is harder to capture with resolved Gaussians using \textsc{Aegean-Tools}. For such sources, we ignore the contribution by the diffuse emission, assuming that the detected components at the nominal SNR$>10$ level at both frequencies dominate the energetics given that the uncaptured emission is much lower in surface brightness and in filling factor. Furthermore, such emission is not observed in the corresponding higher frequency map, and for sources with two VLBA maps, we only consider components identified in both maps. There are also sources like J0405+3803 \citep[which has a second core as identified by][]{2004ApJ...602..123M} and J0741+2706, where at the lower S-band frequencies, cores are merged into brighter components (C1 for both sources). The spectral indices of these components, however, suggest that these cores are likely faint at the frequencies imaged.

For each non--core component, we use its reference luminosity $L_0$ corresponding to the rest--frame frequency of the low--frequency map, modeling the intrinsic synchrotron spectrum as a power law (Equation \ref{eq:lum_prof}) and the corresponding component volume from the lower frequency map to calculate the minimum energy magnetic field strength using Equation \ref{eq:b_min} and the minimum energy using Equation \ref{eq:e_min}. We find typical component magnetic field strengths of $\sim 20$ mG and a rather broad range of minimum energies on the order of $10^{-2}$--$10^{2}\,\mathrm{M_{\odot}c^2}$.
%For components with steep spectral indices $\alpha\leq-2$ that are suspected to be due to resolved-out emission, we compare to existing spectral index maps in literature. We identify C3 of J0405+3803 \citep{2004ApJ...602..123M}, C2 of J0741+2706 \citep{2016MNRAS.459..820T}, 

Cores are filtered out in our calculations by their generally flat or inverted spectral indices $\alpha>-0.5$; we also confirm by comparing to existing literature on these sources. The following sources had a detected (SNR $>$ 10) core at two frequencies that was filtered out:  J0943+1702, J1644+2536, J1234+4753, J1244+4048 \citep{2016MNRAS.459..820T}, J1511+0518 \citep{2012ApJS..198....5A}, J1158+2450 \citep{2017MNRAS.471.1873Y}, J1945+7055 \citep{1999ApJ...521..103P}, and J2022+6136 \citep{2000A&A...360..887T}. 
%\begin{figure}[htp!]
%    \centering
%    \includegraphics[width=\linewidth]{imgs/ind_hotspot_res.pdf}
%    \caption{Distributions of magnetic field strengths and minimum energies over individual components in the CSO-2 sample with detected cores removed. Medians are labeled and marked with the red vertical dashed line.}
%    \label{fig:ind_hotspots}
%\end{figure}

\subsection{Spectral Analysis}\label{sec:spec_anal}
A significant concern with using VLBA map--fitted component fluxes is the possibility of resolving out flux compared to single-dish or short-baseline interferometric observations. We therefore consider using the broadband spectrum to constrain our estimates of the overall synchrotron luminosity of each source, avoiding VLBA total flux densities. We use a double power law model that enables a fit of the optically thin and optically thick spectral index:
\begin{equation}
    F_{\nu,\text{obs}}(\nu) = \frac{F_0}{1-e^{-1}}\left(\frac{\nu}{\nu_0}\right)^{\alpha_1}\left(1-e^{\left(-\frac{\nu}{\nu_0}\right)^{\alpha_2-\alpha_1}}\right).
\label{eq:prof}
\end{equation}
Here, $F_0$ and $\nu_0$ correspond to a reference flux density and frequency, $\alpha_1$ is the optically thick spectral index (which can be $<0$ if not fully absorbed) and $\alpha_2<0$ is the optically thin spectral index \citep{1998A&AS..131..435S}. We note that $F_0$ and $\nu_0$ are not exactly at the break of the spectrum, as stated by \citet{1998A&AS..131..435S}, but will be close in practice; the break frequency and flux density are not analytically found. An alternative parameterization is to do a parabola in log-space, allowing for an estimate of the observed luminosity without knowledge of $\gamma_{\min}$, $\gamma_{\max}$, and the magnetic field. However, we find that it tends to underestimate the highest (lowest) frequency measurements, which are more consistent with having minimal spectral curvature with respect to other observations in the optically thin (thick) portions of the spectrum closer to the turnover.

\begin{figure}
    \centering
    \includegraphics[width=\linewidth]{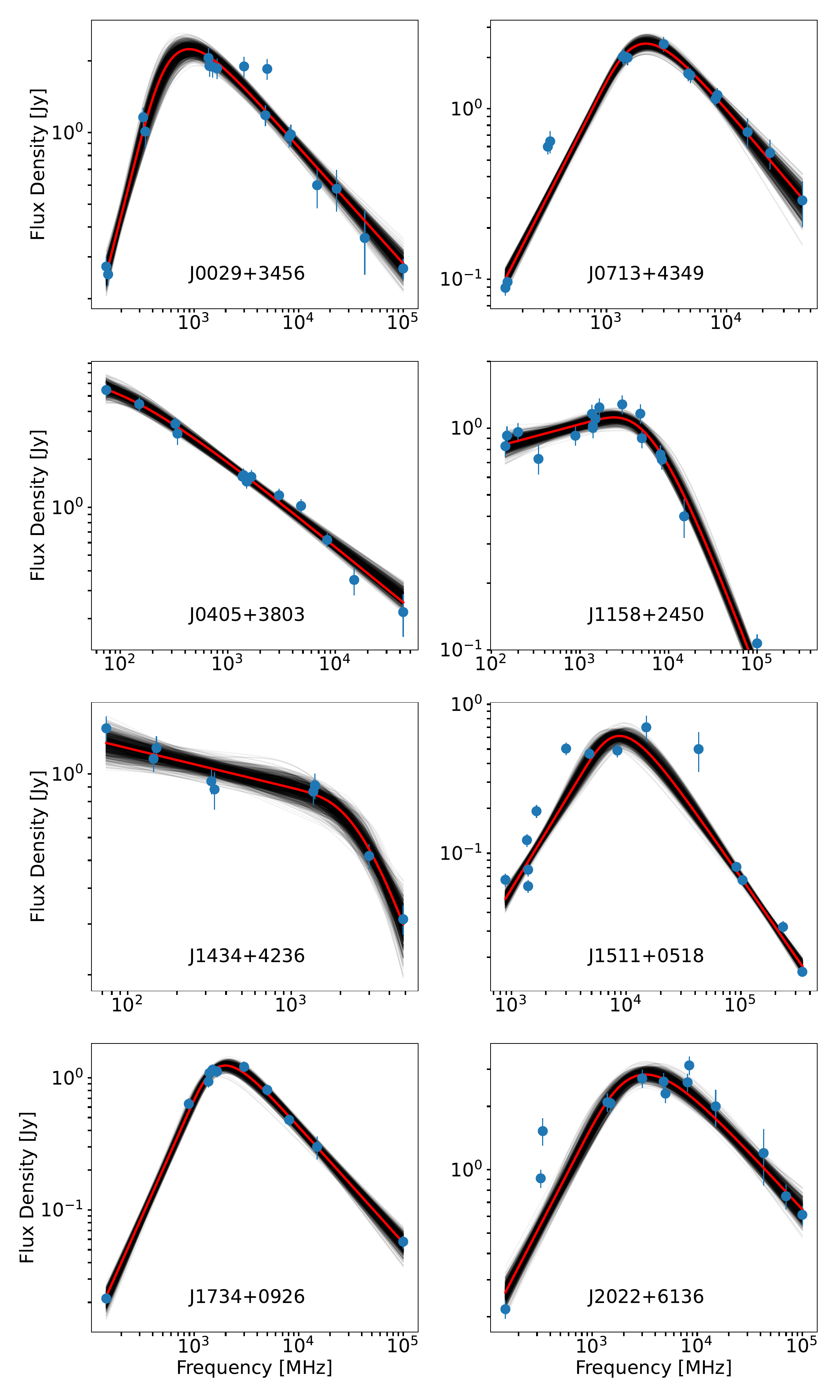}
    \caption{Selection of CSO-2 radio spectra, showing flux density measurements (blue) and probabilistic double power law fits. In particular, we overlay the data with 1000 posterior samples (black) and the MAP (red).}
    %Sample double power law fits overlaid with the MAP (red) and 1000 posterior draws (black).
    \label{fig:spec_fits1}
\end{figure}

Sample posterior draws along with the maximum a posteriori (MAP) are shown in Figure \ref{fig:spec_fits1}. Fits are done using an MCMC in order to obtain more robust uncertainties on the model parameters, using wide uniform priors on all parameters and requiring the reference flux and frequency to be within the range of the data, with the exception of J0405+3803, which does not appear to have any break frequency over the observed range (see Figure \ref{fig:spec_fits1}). In this case, we set the upper bound on the prior to the break frequency to be the lowest frequency measurement. We allow the optically thick spectral index to be negative for cases like J1434+4236 that are more weakly absorbed (see Figure \ref{fig:spec_fits1}) and may only see a turnover well below the frequencies it has been observed at. For J0713+4349, we narrow the range of the break frequency to be above 2 GHz to obtain a reasonable fit to the optically-thin regime (also in Figure \ref{fig:spec_fits1}). Trace plots of the MCMC chains were visually inspected to confirm convergence in the fit parameters. The ACC 100 GHz measurements are able to strongly constrain our estimate of the optically thin spectral index and their general consistency with lower frequency measurements suggest that the global synchrotron cutoff may occur above 100 GHz. However, such measurements are not available in all cases. Significant deviations from the double power law fit may be due to the intrinsic variability of the sources (particularly the cores), given that data was taken from surveys taken at different times ($>$20 year time range), and calibration errors. In particular, we find that for J0713+4349, J1511+0518 and J2022+6136, which are also X-ray sources, more than one radio flux density measurement exists that may be inconsistent with a strict double power-law. We show the radio spectra of these CSO-2s in Figure~\ref{fig:spec_fits1}. The high-frequency data of J0713+4349 and J2022+6136 are consistent with a constant optically thin spectral index, so we do not expect the fit to bias the energy and magnetic field calculations. For J1511+0518, the high-frequency measurements from the ACC also help solidify the high-frequency spectral index despite the potentially discrepant VLA CC measurements at 22 and 43 GHz. We confirm the result of \cite{2024ApJ...977..195D} that known CSO-2s exhibit a broad range of spectral types, although the majority have a steep high-frequency spectrum: $\alpha_2<-0.5$. 

For the volumes, we use the sum of the lower-frequency map components (and core, if detected) derived in Section \ref{sec:ind} from the RFC VLBA map. Since we only require one VLBA map, we include the 9 additional sources that did not have two VLBA maps at two frequencies into this analysis, applying the same method to fit the component volumes (see Figure \ref{fig:component_fit} and Table \ref{tab:rfc_meta} for the specific RFC maps used). We use the reference flux density at 50 GHz, well into the optically thin regime, and the optically thin spectral index to parameterize the intrinsic luminosity density profile with Equation \ref{eq:lum_prof}, correcting for redshift. We again use Equations \ref{eq:b_min} and \ref{eq:e_min} to calculate the magnetic field strengths and minimum energies. For sources with identifiable cores, we do not consider subtracting the core as its contribution to the radio spectrum is not well understood from the typical two RFC VLBA maps at different frequencies. We note that the resulting magnetic fields are then an average over all components of the source (including the core). Including the core, nevertheless, provides a better estimate of the energy lower bound for tests of the CSO-TDE scenario.

\section{Minimum Energy Analysis Results}\label{sec:min_e_res}

In Table \ref{tab:ind_results}, we report the intrinsic component sizes, fluxes, spectral indices, and minimum energy quantities from the individual component analysis. In Table \ref{tab:spec_results} we list the spectral indices and minimum energy quantities from the spectral analysis (statistical errors only). \citetalias{2024ApJ...961..242R} note that, owing to their asymmetry, the sources J1159+5820, J1227+3635, J1244+4048, and J1400+6120 are likely mildly relativistically boosted and were thus excluded from their analysis. Indeed, we find that all of these sources have minimum energies generally larger than the rest of the sample with $E_{\min}\gtrsim50\,\mathrm{M_{\odot}c^2}$, and are flagged in the tables. However, for mildly relativistic lobe speeds of $\beta\approx 0.4$ observed in CSOs and $L_{\nu}\propto f_{\nu}\propto D^{3-\alpha}$ for Doppler factor $D$, given that $E_{\min}\propto L_{\nu}^{2/(3-\alpha)}\propto D^2$, jet viewing angles of $0^{\circ}\leq \theta\leq 45^{\circ}$ produce $1.28\leq D\leq 1.53$, which can lead to $\leq 2.3\times$ larger energies than estimated, comparable to the level of systematics that exist in the modeling as discussed later in this section. We consider the minimum energy and magnetic field strength statistics of the CSO-2 sample by ignoring these four sources. However, we also provide a separate statement of the statistics if these sources were included in Section \ref{sec:systems}.

\begin{figure}[t!]
    \centering
    \includegraphics[width=\linewidth]{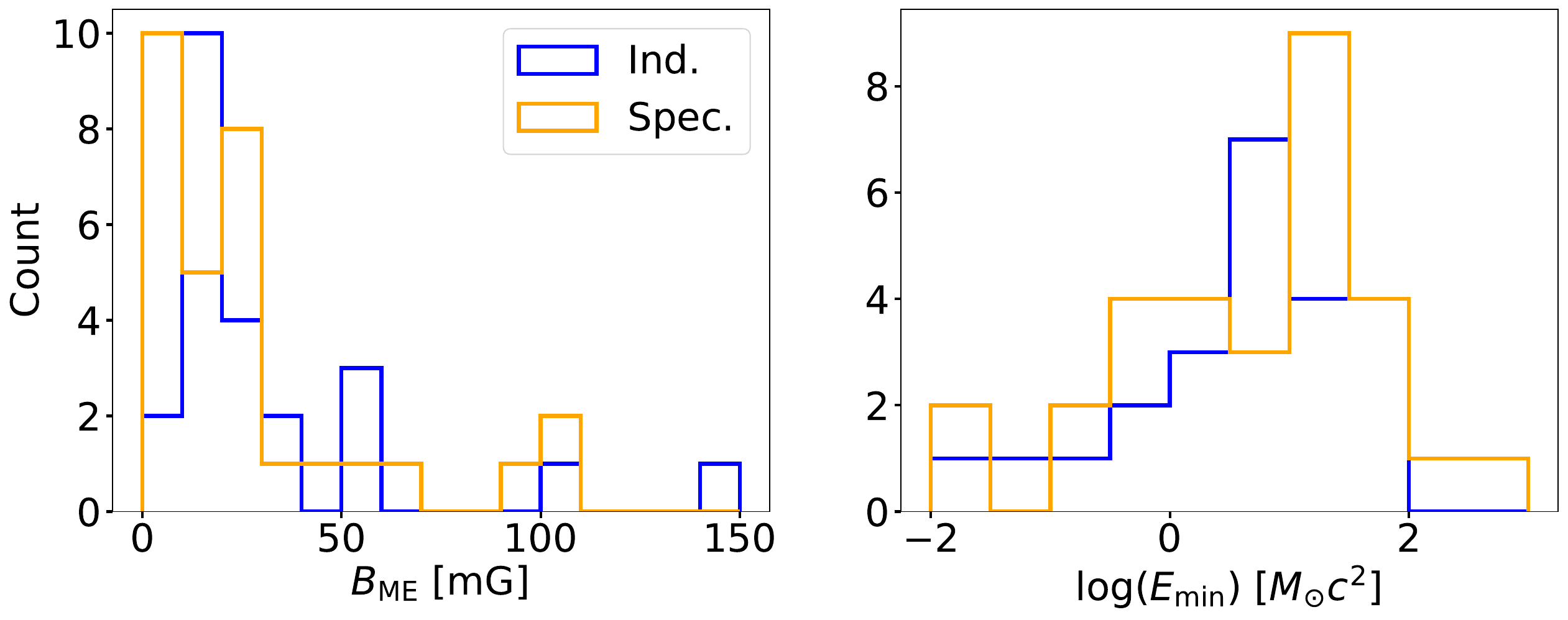}
     \caption{Distribution of averaged minimum-energy magnetic field strengths $B_{\mathrm{ME}}$ (left) and summed component minimum energies $E_{\min}$ (right) from the individual component analysis (blue) and spectral analysis (orange).}
    \label{fig:comp1}
\end{figure}

Ignoring the potential four beamed sources, we show a histogram distribution of posterior median magnetic fields and total minimum energies for both methods in Figure \ref{fig:comp1}. By resampling the individual posterior distributions, we can report uncertainties on the distribution median. Under our fiducial assumptions, the median magnetic field strengths resulting from the individual component analysis and the spectral analysis are $19.7^{+0.8}_{-0.4}$ mG and $19^{+2}_{-2}$ mG, respectively. The median total minimum energies are $5.0^{+0.2}_{-0.4}$ $\mathrm{M_{\odot}c^2}$ and $9.5^{+0.6}_{-0.5}$ $\mathrm{M_{\odot} c^2}$, respectively. However, the means of the energy distributions are more consistent than the medians across the two methods when taken in log-space, with $3.6^{+0.1}_{-0.1}$ $\mathrm{M_{\odot}c^2}$ and $4.4^{+0.2}_{-0.2}$ $\mathrm{M_{\odot}c^2}$, respectively. 

\begin{figure}[t!]
    \centering
    \includegraphics[width=\linewidth]{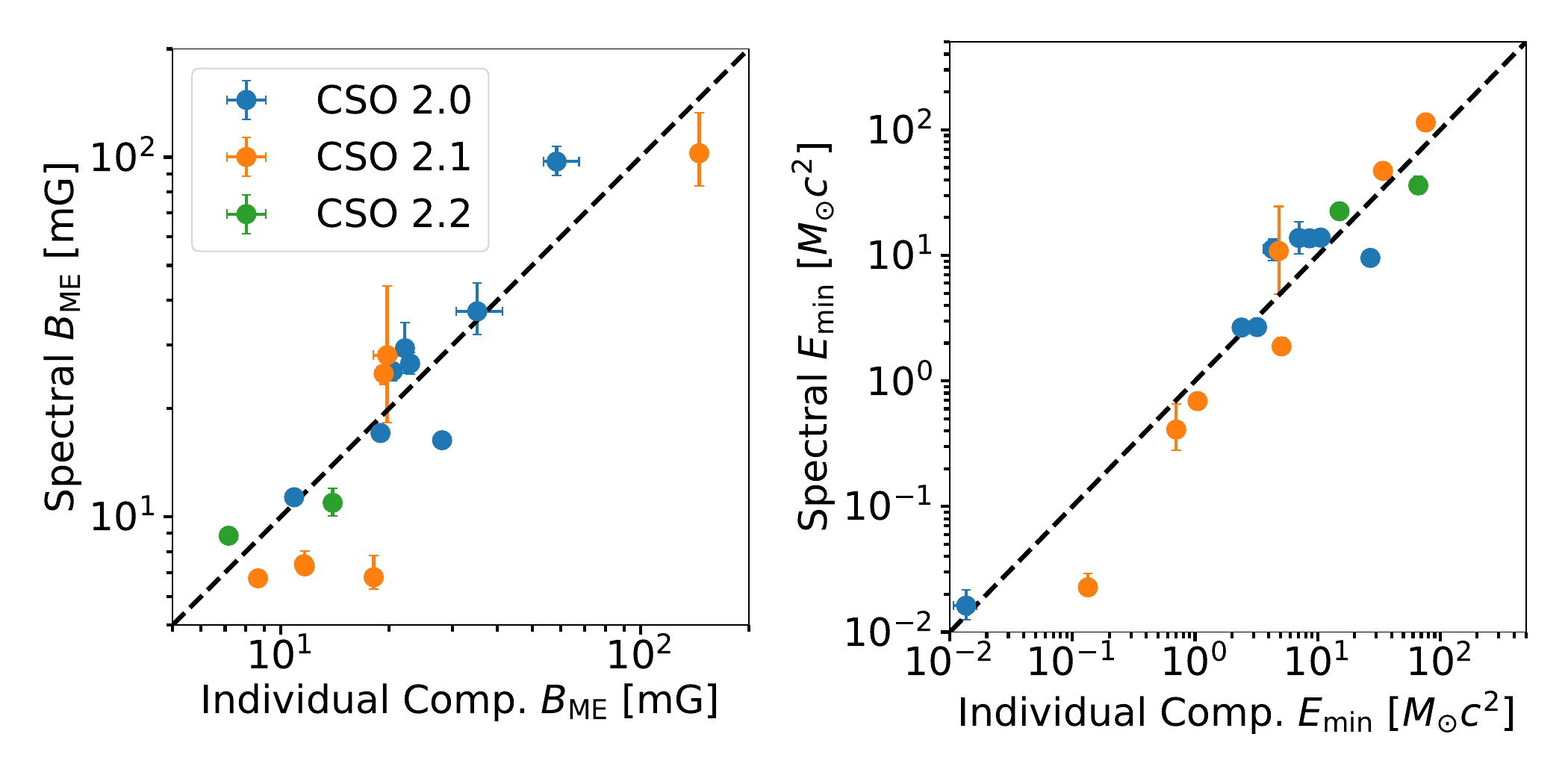}
    \caption{Source-by-source comparison of minimum energy magnetic field strengths (left) and minimum energies (right) across the two methods colored by the CSO-2 subclassification. For each source, we average across component magnetic field strengths and sum the component minimum energies from the individual component analysis.}
    \label{fig:comp2}
\end{figure}

\begin{figure}[ht!]
    \centering
    \includegraphics[width=\linewidth]{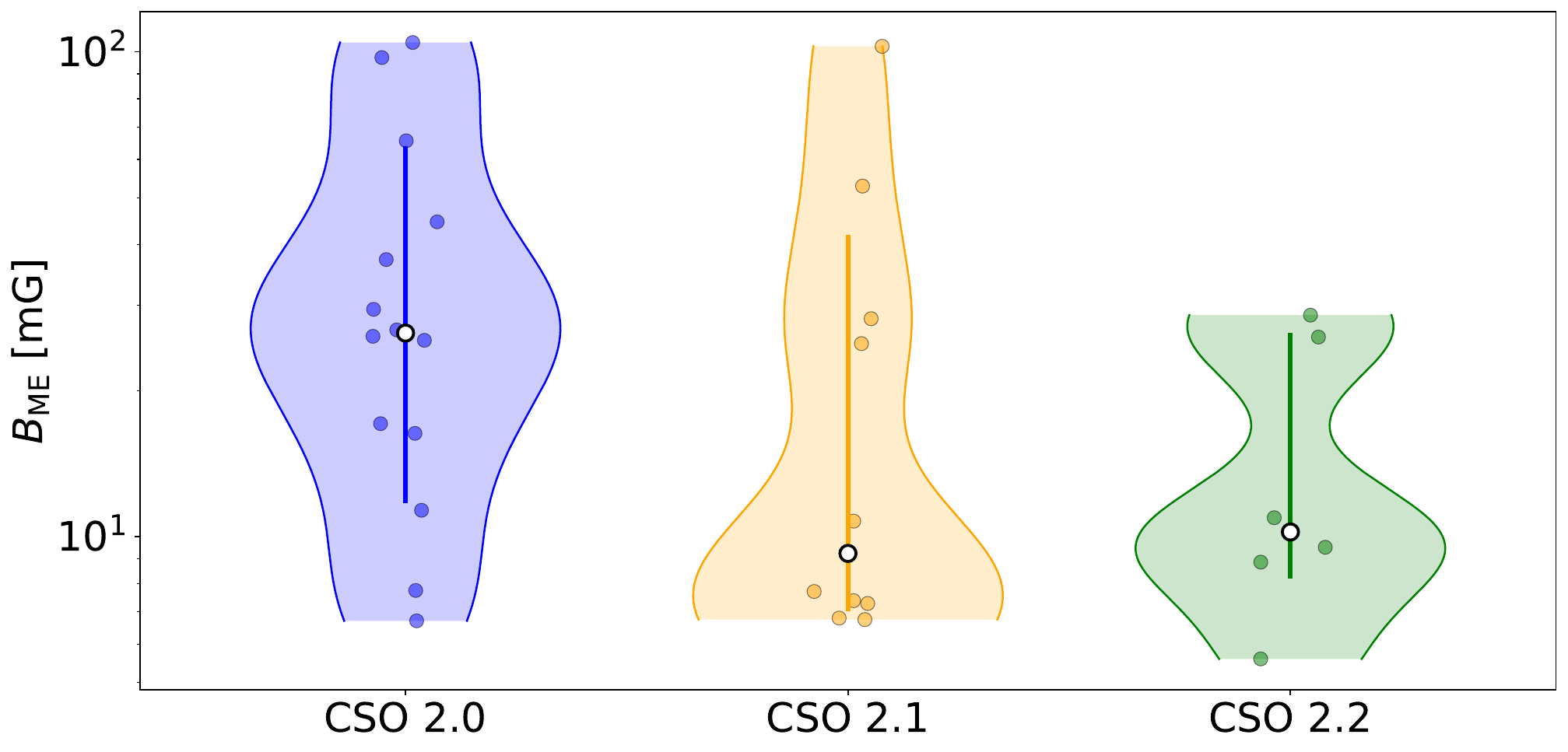}
    \caption{Violin plot of the minimum energy magnetic field strengths from the spectral method by CSO-2 subtype with the 16--84\% interval shown. While the sample size is limited, there may be evidence that the CSO 2.1 and 2.2s have weaker magnetic field strengths than CSO 2.0s, suggesting that they are more evolved.}
    \label{fig:comp3}
\end{figure}

In Figure \ref{fig:comp2}, we compare the minimum energy quantities for sources without cores across the two methods colored by CSO-2 subclasss. In about two-thirds of the cases the magnetic field strengths derived from these two approaches are in reasonable agreement, while in the remaining one-third the spectral analysis underestimates the magnetic field strength relative to the individual component analysis. The minimum energies appear to be more consistent to within a factor of a few. We identify 18/23 from the individual component analysis and 22/30 CSOs from the spectral analysis to have minimum energies larger than $\mathrm{M_{\odot}c^2}$. We note that 16/18 of those from the individual component analysis with energies $>1\,\mathrm{M_{\odot}c^2}$ are in common with the spectral analysis, while the remaining two have minimum energies above and below $\mathrm{M_{\odot}c^2}$ but consistent to within a factor of ${\sim}2$. No clear trend appears in the energetics of the CSO-2 subclasses, but we find that the CSO-2.2s and most the 2.1s tend to have weaker magnetic field strengths than the 2.0s, consistent with 2.2s being more evolved. In Figure \ref{fig:comp3}, we plot the distribution of magnetic field strengths from the spectral analysis for each CSO-2 subtype, generally supporting this observation when extending the larger sample of CSO-2s with cores. We find no distinction in the fraction of sources that exhibits minimum energies $>1\,\mathrm{M_{\odot}c^2}$ between the CSO-2 subtypes. Namely, the fraction of CSO 2.0s, 2.1s, and 2.2s in the sample of 30 used for the spectral analysis are consistent with the fraction of those that are $>1\,\mathrm{M_{\odot}c^2}$.

\subsection{Potential Systematics}
\label{sec:systems}

\begin{deluxetable*}{lcccc}[htp!]
\tablecaption{Summary of Systematics Tests on the Median Minimum Energy Magnetic Field Strengths $B_{\mathrm{ME}}$ and Minimum Energies $E_{\min}$ on the Individual Component Analysis (Subscript 1) and Spectral Analysis (Subscript 2)\label{tab:tests}}
\tablehead{
\colhead{Test} & 
\colhead{$B_{\mathrm{ME}, 1}$ (mG)} & 
\colhead{$B_{\mathrm{ME}, 2}$ (mG)} & 
\colhead{$E_{\min, 1}$ ($M_{\odot}c^2$)} & 
\colhead{$E_{\min, 2}$ ($M_{\odot}c^2$)}
}
\startdata
$10\leq \gamma \leq 10^{5}$ (fiducial)  & $19.7^{+0.8}_{-0.3}$ & $19^{+2}_{-2}$ & $5.0^{+0.2}_{-0.4}$ & $9.5^{+0.6}_{-0.5}$ \\
$10\leq \gamma \leq 10^{4}$  & $19.4^{+0.5}_{-0.2}$ & $18.8^{+1.3}_{-2.5}$ & $5.0^{+0.2}_{-0.4}$ & $8.0^{+0.4}_{-1.3}$ \\
$10 \leq \gamma \leq 10^{3}$ & $19.2^{+0.1}_{-0.3}$ & $17.8^{+1.1}_{-1.9}$ & $5.0^{+0.2}_{-0.4}$ & $6.5^{+0.5}_{-0.9}$ \\
$1\leq \gamma \leq 10^{5}$   & $34.2^{+0.5}_{-0.5}$ & $26.5^{+1.5}_{-1.6}$ & $16.5^{+0.4}_{-0.5}$ & $13.4^{+1.9}_{-1.8}$ \\
$50\leq \gamma \leq 3000$    & $12.9^{+0.4}_{-0.1}$ & $12.7^{+0.4}_{-0.3}$ & $2.1^{+0.2}_{-0.2}$ & $3.7^{+0.8}_{-0.6}$ \\
Excl. Broad X-ray Emission   & $18.9^{+0.2}_{-0.5}$ & $13.9^{+0.4}_{-0.5}$ & $5.4^{+1.5}_{-0.3}$ & $10.3^{+1.2}_{-0.8}$ \\
Excl. Steep Spectrum         & $19.5^{+0.4}_{-0.4}$ & $10.5^{+0.5}_{-0.5}$ & $6.1^{+0.2}_{-0.1}$ & $9.2^{+0.3}_{-1.2}$ \\
Excl. VLBA $>2$ Years        & $20.7^{+1.2}_{-0.4}$ & \nodata              & $4.1^{+0.5}_{-0.6}$ & \nodata              \\
Incl. Asymmetric Sources & $19.4^{+0.5}_{-0.2}$ & $15.9^{+0.6}_{-0.7}$ & $5.3^{+1.5}_{-0.3}$ & $10.3^{+1.4}_{-1.1}$
\enddata
\tablecomments{The last four rows assume the fiducial $10\leq\gamma\leq 10^{5}$. All tests ignore the four CSO-2s that are potentially mildly relativistically boosted, with the exception of the last one, where they are included.}
\end{deluxetable*}

The discrepancies between methods are likely due to the following systematics: (i) the resolving out of flux in the VLBA maps, and (ii) the averaging of the spectral index across individual components in the spectral analysis. (i) can lead to systematically higher component-wise magnetic field strengths and energies due to steeper spectral indices, particularly for those that are low-surface brightness and large volume. Several individual components in Table \ref{tab:ind_results} have spectral indices $\alpha < -1.5$, which make up most of the component-wise magnetic field strength measurements $\gtrsim75\,\mathrm{mG}$. These magnetic field estimates and corresponding energies are likely overestimated and should thus be interpreted with caution. (ii) can have the opposite effect, dampening out steep spectral indices from large volume, low surface brightness components that have large energies or magnetic field strengths. In the spectral analysis, the brightest components (which may include the core) will set the observed spectral index, potentially leading to milder spectral indices compared to the individual components. While flatter spectral indices typically lead to lower magnetic field strengths, better coverage of the overall synchrotron luminosity constrained by the broadband radio spectrum typically estimates larger energies.

Another potential source of systematic error is the choice of electron $\gamma_{\min}$ and $\gamma_{\max}$. We find that decreasing $\gamma_{\max}$ to $10^{4}$ or $10^{3}$ does not significantly impact the individual component analysis results; all distribution medians are generally consistent within the error bars. However, for the spectral analysis, we find a small reduction in the magnetic field strengths and minimum energies by ${\sim}0.9\times$ in the magnetic field strength and ${\sim}0.7\times$ going down to $\gamma_{\max}=10^{3}$. The highest frequencies contribute the least to the total synchrotron luminosity and fewer electrons have high enough $\gamma$ to emit at these frequencies. Consequently, increasing $\gamma_{\max}>10^{5}$ will also not significantly impact the minimum energy quantities. We conclude that spectral aging of high energy electrons would not significantly change our conclusions.

On the other side, by decreasing $\gamma_{\min}$ to 1, we find the median magnetic field strength increases by ${\sim}1.7\times$ and ${\sim}1.3\times$ from the individual component analysis and spectral analysis, respectively. The median minimum energy also increases by a factor of ${\sim}3.3\times$ and ${\sim}1.4\times$, respectively. This is because there would be more electrons emitting at a lower energy (or $\gamma$), and the the lowest frequencies dominate the intrinsic synchrotron luminosity density under an extrapolation of the optically thin spectral index. Finally, considering the observable range of frequencies (54 MHz to 442500 MHz), for typical magnetic field strengths on the order of $10\,\mathrm{mG}$, we restrict the electron energy range from $\gamma_{\min}\approx50$ to $\gamma_{\max}\approx3000$. The median magnetic field strengths become $12.9^{+0.4}_{-0.1}\,\mathrm{mG}$ and $12.7^{+0.4}_{-0.3}\,\mathrm{mG}$ in the individual component and spectral analyses, respectively. The minimum energies drop to $2.1^{+0.2}_{-0.2}\,\mathrm{M_{\odot}}c^2$ and $3.7^{+0.8}_{-0.6}\,\mathrm{M_{\odot}}c^2$ respectively, but these frequency bounds are likely not representative of the intrinsic synchrotron spectrum of the CSO-2s.

We investigate the impact on the median magnetic field strengths and energies relative to the fiducial case caused by the inclusion or exclusion of sources that may introduce systematic effects. We still assume the fiducial $10\leq\gamma\leq 10^{5}$. Removing the X-ray sources that show evidence of broad X-ray iron emission (three of which show potential deviations from the radio double power law fits), changes the magnetic field strengths by ${\sim}0.95\times$ and ${\sim}0.7\times$ and minimum energies by ${\sim}1.1\times$ and ${\sim}0.9\times$ across the individual component and spectral analyses, respectively. Alternatively, removing the sources that contain at least one steep-spectrum component with $\alpha<-1.5$ in the individual component analysis and global spectrum with average $\alpha<-1$ in the spectral analysis, we find changes of ${\sim}1\times$ and ${\sim}0.55\times$ in the magnetic field strengths and ${\sim}1.2\times$ and ${\sim}0.97\times$ in the minimum energies across the two methods respectively. Ignoring the seven sources with VLBA maps separated by more than two years, we find changes of ${\sim}1.05\times$ in the magnetic field strengths and a ${\sim}0.8\times$ change in the minimum energy for the individual component analysis. We conclude that the median minimum energies are generally robust to the systematics created by the inclusion of sources with unusually steep spectrum components and potential core variability. Finally, with the inclusion of the four asymmetric sources, we find that the median minimum energies are consistent within the errors of the fiducial case. These various systematics tests are summarized in Table \ref{tab:tests}. In any case, our conclusion that the majority of the VLBA--detected CSO-2s are consistent with having minimum energies $>1\,\mathrm{M_{\odot}}c^2$ is unchanged.

%Accurately modeling the location of cooling cutoffs in synchrotron spectra for both the entire source and individual components remains a challenging problem. There is little evidence that a strong cutoff exists down to ${\sim}100\,\mathrm{GHz}$ frequencies in the overall synchrotron spectra. The cooling times for $\sim100 \,\mathrm{GHz}$ radiating particles should be far smaller than the CSO-2 kinematic timescale.

\begin{figure}[t!]
    \centering
    \includegraphics[width=\linewidth]{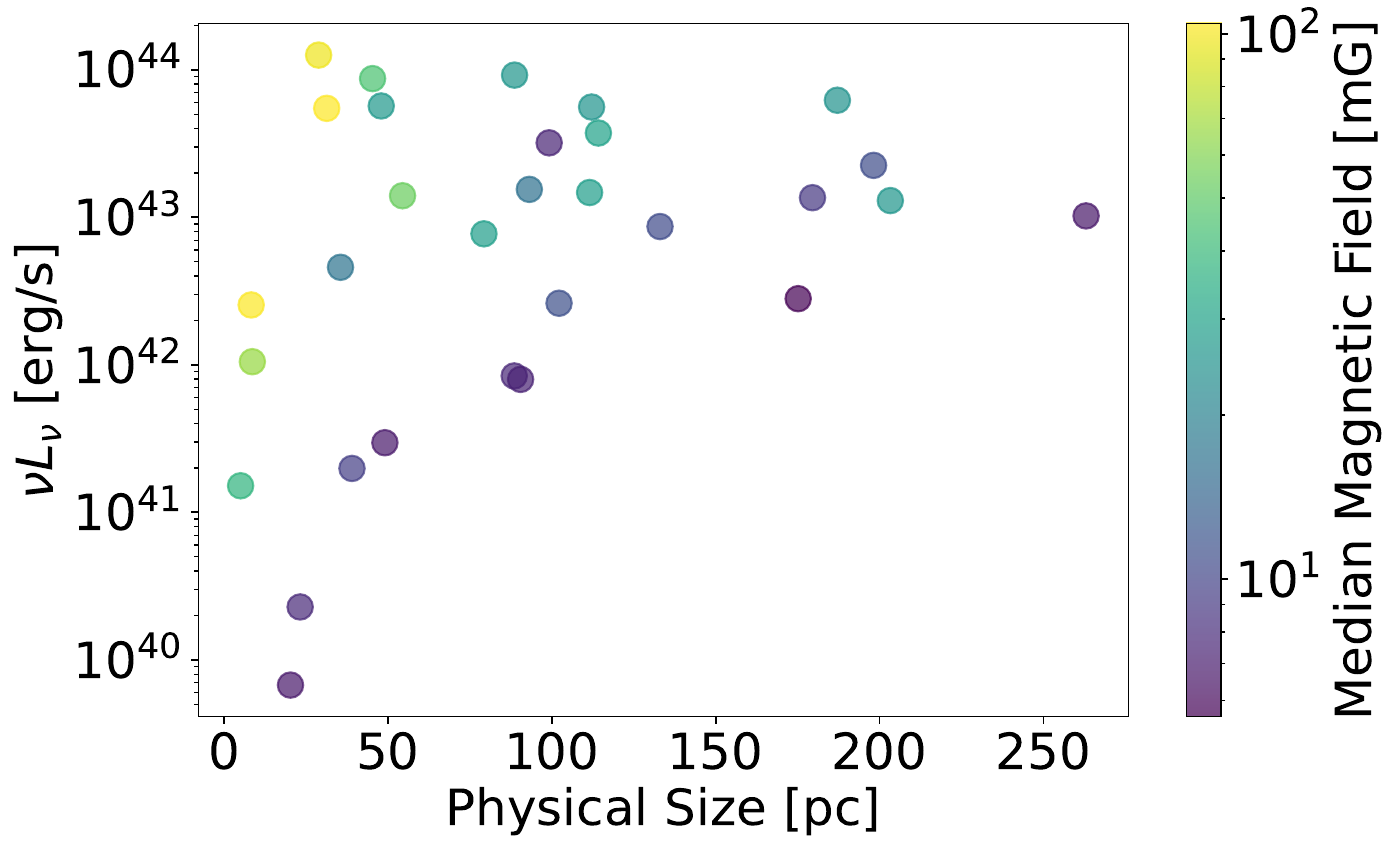}
    \caption{Minimum energy magnetic field strengths as a function of the 
    \textit{approximate} peak luminosity (derived using $F_0$ and $\nu_0$ from Equation \ref{eq:prof}) and half the projected length of the CSO-2s analyzed with the spectral analysis method.
    %physical size-->length
    We have taken the median of the posterior distribution in the magnetic field strengths and luminosities.}
    \label{fig:field_summ}
\end{figure}

Finally, we note that these minimum energies and magnetic field strengths are determined for the brightest CSO-2s in the sky and are therefore expected to be the largest among the CSO-2s for any given physical size, as shown in Figure \ref{fig:field_summ}. While we expect that in the typical size regime of CSOs that the radio luminosity should gradually rise \citep{10.1093/mnras/stx3358}, the radio source size will likely grow faster, resulting in an overall decrease in radio surface brightness. Hence, conservation of magnetic flux requires that at large physical sizes, magnetic field strengths should decrease despite the surface brightness selection effect, which is also observed in Figure \ref{fig:field_summ}. We expect a large population of CSO-2s at intermediate sizes not detectable within VLBA surface brightness limits with significantly lower luminosities and weaker magnetic field strengths.

\section{Proximity to Minimum Energy}
\label{sec:inv_comp}
In order to better estimate the energetics and cooling times of CSO-2s, we seek estimates of the true magnetic field strengths of the CSO-2s in the sample, rather than the minimum energy magnetic field strengths. There is strong evidence that CSOs and Gigahertz-Peaked Spectrum (GPS) sources are close to minimum energy \citep{1994ApJ...426...51R, 2003PASA...20...69P, 2008A&A...487..885O}. Methods for determining how close radio sources are to minimum energy come in two forms: (i) under the assumption that the source is synchrotron self-absorbed, the spectral turnover can directly constrain  estimates of the magnetic field \citep{1977MNRAS.180..539S,2008A&A...487..885O}, and (ii) modeling the inverse Compton X-ray emission from CMB, AGN, or synchrotron photons (synchrotron self-Compton, SSC) generated by the same electrons emitting synchrotron \citep{2005ApJ...626..733C, 2008ApJ...680..911S, 2010ApJ...715.1071O}. In this section, we focus on method (ii) to examine whether comptonization by the electron population in the components can explain the observed X-ray emission identified in a subset of CSO-2s. We note that method (i) may be less reliable given that it is often difficult to disentangle free-free and synchrotron self-absorption, and both may generally contribute to the break in the spectrum \citep{Marr_2014, Tingay_2003, 2024ApJ...977..195D}. Furthermore, if self-absorption is the case, an upper bound on the magnetic field can only be obtained if the angular size at spectral turnover is known \citep[see updated formula in Appendix B of][]{Readhead_2021}, which is typically only approximately determined.

\begin{figure*}[htp!]
    \centering
    \includegraphics[width=\linewidth]{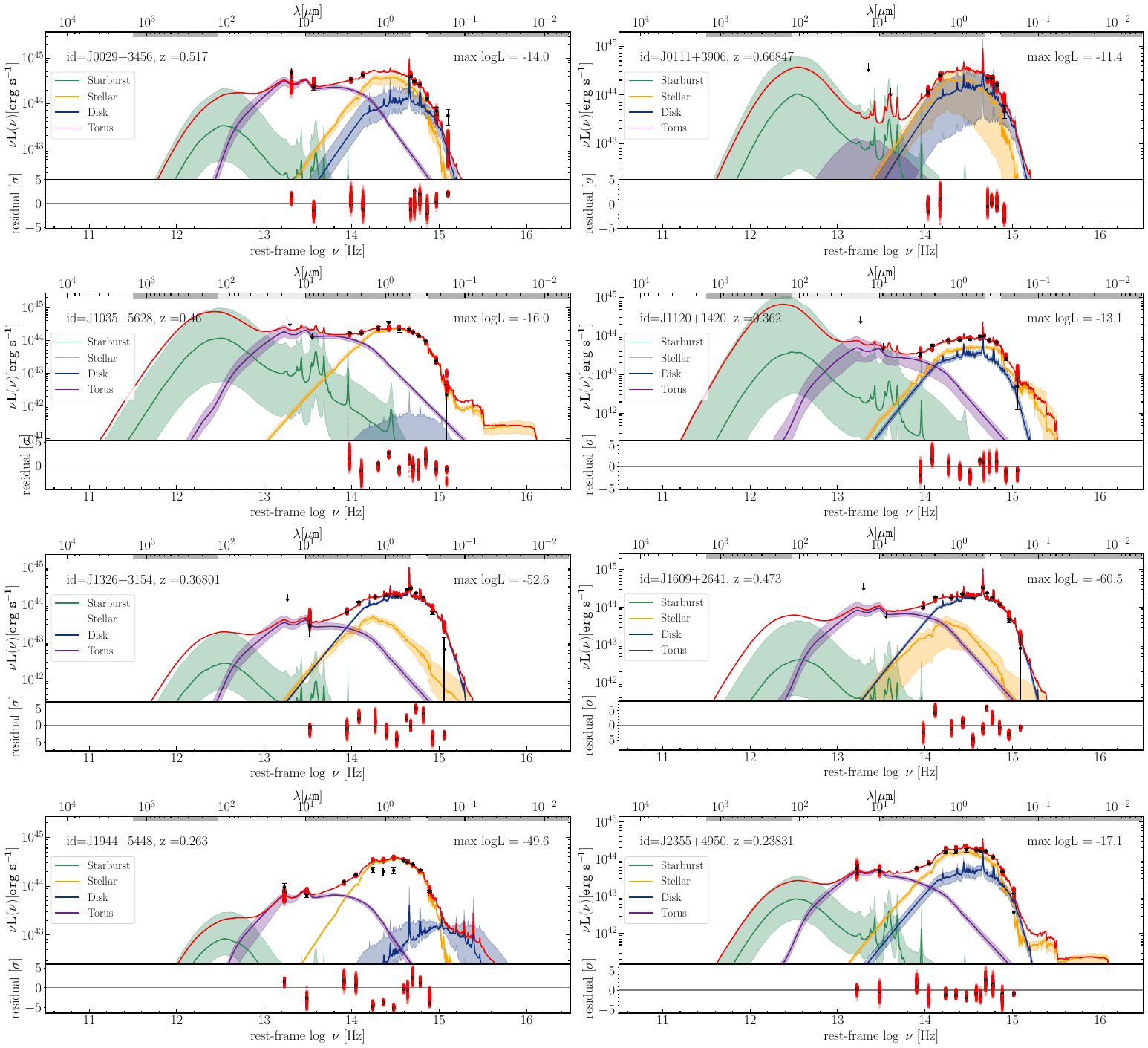}
    \caption{\textsc{AGNFitter} SEDs fitted to the IR--UV flux densities of the host galaxies of 8 CSO-2s analyzed for X-ray inverse Compton emission.
    The shaded regions correspond to the 1$\sigma$ intervals from the posterior, with the solid line representing the median.
    Degeneracies between the starlight \textit{(yellow)} and reddened AGN accretion disk \textit{(blue)} components are evident.
    Ameliorating this degeneracy will be the subject of future work.}
    %and improved resolution of this degeneracy will be subject to future work.}
    \label{fig:agnfitter}
\end{figure*}

With the ability of \textit{Chandra} to resolve X-ray emission confined to the radio lobes of FR IIs, \cite{2005ApJ...626..733C} showed that comptonization of CMB photons could explain their X-ray fluxes, yielding magnetic field strengths that are within order unity of equipartition and thus minimum energy. We note that inverse Compton of non-CMB photon fields have also been considered to account for X-ray fluxes in such radio galaxies \citep[e.g., see][and references within]{2001A&A...372..755B, 2008MNRAS.386.1774E, 2023A&A...672A.179B}. A similar analysis for CSO-2s is particularly challenging; given the largest separation of ${\sim}500$ pc (corresponding to sub-arcsecond scales), the radio lobes are unresolved in \textit{Chandra}, and comptonization is expected to occur of host galaxy, AGN, and synchrotron emission rather than primarily of CMB photons, given the formers' larger energy density. \cite{2008ApJ...680..911S, 2010ApJ...715.1071O} argue that the X-ray emission of GPS sources can be well-explained by inverse-Compton scattering of non-CMB radiation fields, thus motivating this analysis.

\begin{deluxetable*}{lcccccc}
\tablecaption{Inverse Compton Analysis Results} \label{tab:ic_results}
\tablehead{
\colhead{Name} &
\colhead{Disk ($\log(L_\odot)$)} &
\colhead{Starlight ($\log(L_\odot)$)} &
\colhead{Torus ($\log(L_\odot)$)} &
\colhead{X-ray ($\log(L_\odot)$)} &
\colhead{$B/B_{\mathrm{ME}}$} &
\colhead{$E/E_{\min}$}}
\startdata
J0029+3456 & $45.5^{+0.3}_{-0.3}$ & $44.35^{+0.08}_{-0.20}$ & $44.84^{+0.05}_{-0.05}$ & $44.13 \pm 0.04$ & $4.7^{+0.7}_{-0.6}$ & $9^{+3}_{-2}$ \\
J0111+3906 & $45.7^{+0.3}_{-0.9}$ & $44.0^{+0.3}_{-0.9}$ & $39^{+4}_{-5}$ & $43.6 \pm 0.2$ & $0.72^{+0.17}_{-0.15}$ & $1.3^{+0.6}_{-0.2}$ \\
J1035+5628 & $41^{+2}_{-2}$ & $44.235^{+0.010}_{-0.009}$ & $44.59^{+0.10}_{-0.08}$ & $43.610 \pm 0.007$ & $0.6^{+0.3}_{-0.2}$ & $1.7^{+2.3}_{-0.6}$ \\
J1120+1420 & $45.04^{+0.07}_{-0.08}$ & $43.73^{+0.04}_{-0.04}$ & $44.0^{+0.2}_{-0.2}$ & $42.8 \pm 0.2$ & $0.74^{+0.18}_{-0.14}$ & $1.18^{+0.37}_{-0.15}$ \\
J1326+3154 & $45.66^{+0.03}_{-0.03}$ & $43.0^{+0.1}_{-0.2}$ & $43.88^{+0.10}_{-0.09}$ & $43.62 \pm 0.07$ & $2.9^{+0.5}_{-0.4}$ & $3.9^{+1.4}_{-1.0}$ \\
J1609+2641 & $45.72^{+0.05}_{-0.03}$ & $42.9^{+0.2}_{-0.3}$ & $44.29^{+0.10}_{-0.08}$ & $43.31 \pm 0.06$ & $0.37^{+0.03}_{-0.02}$ & $4.7^{+0.8}_{-0.7}$ \\
J1944+5448 & $43.8^{+0.6}_{-0.2}$ & $44.413^{+0.008}_{-0.020}$ & $44.23^{+0.03}_{-0.03}$ & $42.7 \pm 0.2$ & $2.7^{+1.5}_{-1.0}$ & $3.3^{+4.4}_{-1.8}$ \\
J2355+4950 & $45.2^{+0.1}_{-0.2}$ & $44.00^{+0.06}_{-0.06}$ & $43.97^{+0.04}_{-0.05}$ & $43.1 \pm 0.1$ & $2.3^{+0.9}_{-0.6}$ & $2.4^{+1.9}_{-0.9}$ \\
\enddata
\tablecomments{Column (1): J2000 name, Column (2): de-reddened disk luminosity (0.1--1 $\mu$m), Column (3): starlight luminosity (0.1--1 $\mu$m), Column (4): torus luminosity (1--30 $\mu$m), Column (5): absorption-corrected rest-frame 2-10 keV X-ray luminosity, Column (6): ratio of the magnetic field required to explain X-ray emission to the minimum energy magnetic field, Column (7): ratio of energies. X-ray luminosities for J0029+3456, J0111+3906, J1326+3154, and J1609+2641 are taken directly from \cite{10.1093/mnras/stae2817}; J1120+1420, J1944+5448, and J2355+4950 are taken from \cite{Gan_2025}, using the standard K-correction $K(z)=(1+z)^{1+\alpha}$ to convert to a luminosity, and J1035+5628 is taken from \cite{2006MNRAS.367..928V}, using the \textsc{Xspec} \texttt{tbabs*ztbabs(powerlaw)} model to correct for Milky Way and host absorption \citep{1996ASPC..101...17A, 1999ascl.soft10005A}.}
\end{deluxetable*}

Using recent X-ray analyses of CSOs done by  \cite{10.1093/mnras/stae2817, Gan_2025}, we focus on a sample of 8 CSO-2s that do not have a detected Fe\,K$\alpha$ line nor a strong radio core emission that would otherwise suggest a significant coronal contribution to the X-ray emission. Both independent analyses suggest that the jet/component radiation likely dominate the X-ray emission of the CSOs, given the characteristic photon spectral indices. Previous inverse-Compton analyses of CSOs \citep{2010ApJ...715.1071O, L_Krol_2024} use black-bodies to model the broadband SED. Here, we use \textsc{AGNfitter} \citep{2016ApJ...833...98C, 2024A&A...688A..46M} to more robustly separate out the host-AGN, dust, and starlight contributions.

We collect IR--UV flux densities from SDSS DR16 \citep{2020ApJS..249....3A}, PS1 DR2 \citep{2020ApJS..251....6M}, 2MASS \citep{2003tmc..book.....C}, and WISE \citep{2014yCat.2328....0C} and additionally from UKIDSS DR9 \citep{2007MNRAS.379.1599L} for those without 2MASS measurements, correcting for galactic foreground extinction \citep{2011ApJ...737..103S}; extinction in W3/W4 is assumed to be negligible. No available FIR emission is found in existing catalog searches for any selected source. Due to possible source confusion, we set the W3/W4 flux for J1326+3154 and W4 flux for J1609+2641 to be an upper limit (despite not given as an upper limit). We use the SDSS model magnitudes and PS1 Kron magnitudes to collect extended emission of the host galaxy. We ignore radio and X-ray emission, considering only fits in the IR--UV range. We use the S17 cold dust, CAT3D dusty torus, BC03 stellar population, and THB21 accretion disk model templates for SED fitting \citep[see Section 3 of][for more details]{2024A&A...688A..46M}. These measurements are not enough to strongly constrain estimates of the cold dust emission in the MIR--FIR, but are sufficient to constrain models of the stellar, torus, and/or disk component. Figure \ref{fig:agnfitter} shows the SED fits for all sources considered, showing that in any case, at least one of the stellar, disk, or torus emission are well-constrained. However, the stellar and disk contributions tend to be strongly degenerate.

Given the lack of knowledge about precise source geometry with respect to the accretion disk and that the components are unresolved in X-ray emission, we make the assumption that the external photon fields are isotropic, and their energy densities are averaged across the maximum linear extent of the source. We ignore the contribution to the IC luminosity by cold dust given that it will be more widely distributed throughout the host galaxy. We approximate the spectral energy density at the location of the components from the AGN emission as
\begin{equation}
    u_{\nu,\text{AGN}} \approx \frac{L_{\nu,\text{AGN}}}{4\pi\left(l_{\mathrm{p}}/2\right)^2c},
\end{equation}
where $l_{\mathrm{p}}$ is the overall projected length of the source, defined by the separation of the symmetric components. Values are taken from Table 3 of \citetalias{2024ApJ...961..240K}.

The spectral energy density due to synchrotron emission is taken to be:
\begin{equation}
    u_{\nu,\text{synch}} \approx \frac{L_{\nu,\text{synch}}}{4\pi V^{2/3}c},
\end{equation}
where $V$ is the total component volume. In general, the synchrotron energy density will exceed the contribution by external photon fields given the smaller volume that the components occupy compared to the overall linear extent ($V^{1/3} < l_{\mathrm{p}}/2$).

Finally, we consider the starlight spectral energy density. Given that the PS1 photometry suggests these CSO-2s are extended sources, we use the package \textsc{imfit} \citep{2015ApJ...799..226E} to fit a  PSF-convolved ($\sim 1^{\prime\prime}$ FWHM) S\'ersic profile to the i-band PS1 cutout images to estimate the S\'ersic index and circularized half-light radius, masking any nearby sources that would bias the fits. We also incorporate a point source for the AGN. PSFs are built using at least two isolated stars near the galaxy with \textsc{ePSF} of \textsc{photutils} \citep{2016ascl.soft09011B}. Because the PS1 cutouts are already background-subtracted, we estimate the sky level as $3\times$ the RMS of the current background above the minimum pixel. This provides half-light radii qualitatively consistent with the images. We approximate the starlight's spectral energy density to be:
\begin{equation}
u_{\nu,\star} \approx \frac{L_{\nu,\star}(r<l_\mathrm{p}/2)}{4\pi(l_\mathrm{p}/2)^2c},
\end{equation}
where $L_{\nu,\star}(r<l_\mathrm{p}/2) = fL_{\nu,\star}$ and $f$ is the fraction of the luminosity within $l_\mathrm{p}/2$, derived using the S\'ersic profile. We note that for typical starlight luminosities of $10^{44}\,\mathrm{erg/s}$, the median torus or disk energy density exceeds the starlight energy density by at least $100\times$, so our conclusions would remain unchanged if the starlight component was not included. However, starlight is still considered for completeness because a few combinations of the SED posterior can result in a non-negligible starlight spectral energy density.

We calculate the inverse Compton volume emissivity using Equation 7.28 in \cite{Rybicki1979}, which holds in the Thomson regime under the assumption that $\gamma\gg1$, emission is isotropic, and there are negligible secondary scatterings. We bootstrap the synchrotron, external photon, and unabsorbed X-ray 2--10 keV rest-frame luminosities to get a distribution over the required magnetic field strength to match the sampled luminosity; given the expensive minimization required to solve the non-analytic equation for the magnetic field, we use only 2500 samples. We have symmetrized the uncertainties on the observed 2--10 keV luminosity and assumed it to be a normal distribution. In Table \ref{tab:ic_results}, we list the host galaxy, torus, and accretion disk (de-reddened) luminosities outputted from \textsc{AGNfitter} as well as the observed 2--10 keV luminosity and the ratios $B/B_{\mathrm{ME}}$ and $E/E_{\min}$, where $B_{\mathrm{ME}}$ is the minimum energy magnetic field and $E_{\min}$ is the minimum energy obtained from the spectral analysis. Averaging over the sources, we find a typical $B/B_{\mathrm{ME}}\sim 1.9$ and a typical $E/E_{\min}\sim 4$, but we note that among individual sources, the statistical uncertainty in the energy ratio is large, with distributions typically having long tails towards large $E/E_{\min}$, and model-dependent uncertainties are not included. Even ignoring J1035+5628 and J1944+5448, which have fractional uncertainties in $B/B_{\mathrm{ME}}$ as large as 50\% and in $E/E_{\min}$ larger than $100\%$, the typical deviations from minimum are unchanged. Three sources are consistent with minimum energy to ${\sim} 1\sigma$ and the rest are consistent to within order unity. These results differ from \cite{2005ApJ...626..733C}, who find a typical $B/B_{\mathrm{eq}}\sim 0.7$ for FR IIs.

We note that the predicted spectral indices from the inverse Compton analysis are not always consistent with the observed X-ray spectral shape, but a direct comparison is likely fraught with uncertainty for some sources with low signal-to-noise X-ray spectra. This requires the photon index to be fixed or often results in large uncertainties in power law fits to observations. Furthermore, photon indices tend to differ significantly between epochs or instruments \citep[e.g., see Table 6 of][]{Gan_2025}. In Table \ref{tab:photon_indices}, we compare the modeled photon index from observations in literature and predicted from the inverse Compton analysis. For sources where the spectral index is not fixed, the agreement is to within $2\sigma$. We additionally note that changes in the magnetic field by a factor of a few can change the predicted X-ray luminosity by an order of magnitude, which already discourages strong deviations from minimum energy. Therefore, while we urge the reader to be cautious about these comparisons, there is evidence that CSO-2s are close to minimum energy or equipartition under the assumption that the 2--10 keV emission is primarily inverse Compton.
\begin{deluxetable}{ccc}
\tablecaption{Photon Indices Modeled from the Observed 2--10 keV X-ray spectra (taken from the same references as mentioned in Table \ref{tab:ic_results}) and from the IC emission. \label{tab:photon_indices}}
\tablehead{
\colhead{Source} & \colhead{$\Gamma_{\mathrm{obs}}$} & \colhead{$\Gamma_{\mathrm{IC}}$}
}
\startdata
J0029+3456 & $1.61^{+0.27}_{-0.24}$ & $1.56^{+0.03}_{-0.03}$ \\
J0111+3906 & $1.80$ (fixed) & $2.22^{+0.09}_{-0.05}$ \\
J1035+5628 & $1.75$ (fixed) & $2.10^{+0.11}_{-0.11}$ \\
J1120+1420 & $1.41^{+0.58}_{-0.74}$ & $1.98^{+0.021}_{-0.021}$ \\
J1326+3154 & $1.65^{+0.19}_{-0.22}$ & $1.70^{+0.025}_{-0.023}$ \\
J1609+2641 & $2.27$ (fixed) & $2.115^{+0.008}_{-0.007}$ \\
J1944+5448 & $1.70$ (fixed) & $1.71^{+0.06}_{-0.05}$ \\
J2355+4950 & $0.75^{+0.44}_{-0.49}$ & $1.61^{+0.05}_{-0.05}$ \\
\enddata
\end{deluxetable}

%Given typical X-ray photon indices of $\sim 1.5$--$2$, $F_{\nu}\propto\nu^{-0.5}$--$\nu^{-1}$, which is consistent with the inverse Compton models of synchrotron Self-Compton
%It is important to note, however, that changes in the magnetic field by a factor of a few is sufficient enough to change the predicted X-ray flux by an order of magnitude.

\section{Discussion}
For characteristic synchrotron frequencies of $1\, \mathrm{GHz}$ for CSO-2s, and typical minimum energy magnetic field strengths of ${\sim}20$ mG, the characteristic single-electron Lorentz factor is $\gamma\sim 100$. Then the typical synchrotron cooling lifetime at VLBA frequencies is given by:
\begin{equation}
    t_{\mathrm{cool}} = (613\,\mathrm{yr})\left(\frac{100}{\gamma}\right)\left(\frac{20\,\mathrm{mG}}{B}\right)^2,
\end{equation}
which would be ${\sim}4\times$ smaller for the average deviation from minimum energy we found in the previous section. The shortest kinematic timescale for the largest CSO-2s is approximately $t_{\mathrm{kin}} \geq l_{\mathrm{p}}/2c \sim 10^{3}\,\mathrm{yr}$. Therefore, if CSO-2s are to abruptly shut off at $l_{\mathrm{p}} \sim 500$ pc in size as suggested by the models of \cite{10.1093/mnras/stae322} and observations of \citetalias{2024ApJ...961..242R}, then CSO-2s will fade on timescales at VLBA frequencies faster than the kinematic timescale. This provides an alternative explanation to prevent the build-up in a population of high-luminosity CSO-2s around 500 pc at VLBA frequencies to those mentioned in \cite{10.1093/mnras/stae322}. Furthermore, the smaller magnetic field strengths in CSO-2.1s and CSO-2.2s corroborate the evolutionary models suggested by \citetalias{2024ApJ...961..242R} and \cite{10.1093/mnras/stae322} that they are more evolved than the CSO-2.0 counterparts.

%Whether the sharp cutoff of CSO-2s at around 500 pc is real remains to be confirmed by deeper surveys of these sources, but it strongly suggests that discrete accretion events can trigger the formation of these sources. It has been suggested by \citetalias{2024ApJ...961..242R} and \cite{10.1093/mnras/stae322} that the stellar capture of an evolved large-radius star $\gtrsim1\, \mathrm{M_{\odot}}$ by a $>10^{8}\, \mathrm{M_{\odot}}$ black hole could explain both the typical kinematic timescales to the CSO-2 maximum size and TDE fallback timescales, assuming that the jet continues to be powered by active accretion.

Among our sample of CSO-2s, which is selected from surveys of the brightest radio sources in the sky, careful modeling of the volume of the emitting regions and the total intrinsic synchrotron luminosity shows that the \textit{typical} minimum energy is close to the maximum $7\,\mathrm{M_{\odot}c^2}$ inferred by \citetalias{2024ApJ...961..242R}. The total energy of the radio--emitting components is likely several factors higher than the minimum energy, accounting for departures from minimum energy, kinetic energy of the bulk motion of the lobes, and work done on the ISM as the CSO-2s expand. Furthermore, additional energy can be accounted for in mass that is yet to be accreted and non--radio accretion disk emission. If energy extraction efficiencies of $<100\%$ are assumed during a TDE, then the CSO-TDE scenario can hold if massive ($> 10\,\mathrm{M_{\odot}}$) evolved stars ($>500\,\mathrm{R_{\odot}}$) are disrupted by massive black holes (see Figure 14 and Equation 4 of \citetalias{2024ApJ...961..242R}) to explain the combination of the fallback timescale, kinematic timescale, and the energies. This may be unlikely given their short lifetimes in the giant phase and typical stellar IMFs. Top-heavy IMFs have been observed in the Milky Way \citep{2010ApJ...708..834B} and inferred in high redshift starforming and local starburst galaxies \citep[e.g.,][and references therein]{2011MNRAS.412..979W, 2024ApJ...970..136G}, but may not be realistic for the elliptical host galaxies of CSO-2s. If the CSO-TDE connection is real, then CSO-2s could provide a window into the IMFs of the cores of elliptical host galaxies. On the other hand, a several solar mass star might require additional energy extraction from the black hole spin to produce these observed energies, which has been invoked previously \citep{1977MNRAS.179..433B, 2022MNRAS.514.5141B}.

Certainly, the CSO-2s observed thus far are likely to be the most luminous and energetic due to selection effects in surveys. On the low energy end, we find that minimum energies of CSO-2s can go down to $\sim 10^{-2}\,\mathrm{M_{\odot}c^2}$, which can more easily be explained by the disruption of less massive stars without invoking high efficiencies. And deeper complete surveys of CSOs might yield a substantial population of lower luminosity CSO-2s compared to those observed thus far with energies $<1\,\mathrm{M_{\odot}c^2}$. But if the high luminosity CSO-2s do turn off after ${\sim}10^{3}$--$10^{4}$ years as the complete surveys suggest, then their origin remains quite uncertain.

Whether these sources can be triggered by other means remains subject to future work. \cite{10.1093/mnras/stae322} note that O($10\,\mathrm{M_{\odot}}$) gas accretion can result in accretion timescales comparable to the lifetimes of CSO-2s, although magnetic pressure or turbulent motion might suspend these gas clouds against collapse. Furthermore, detailed forward modeling is required to check whether the occurrence rates/birth rates of CSO-2s can be commensurate with jetted TDE rates if triggered by more massive, evolved stars than previously thought.
 
\section{Conclusions}
We have presented two different methods of estimating the minimum energy conditions of edge-brightened compact radio sources (CSO-2s). We fit components to RFC VLBA maps to determine their volumes and spectral indices, thus enabling us to estimate the energies and magnetic field strengths of individual components for which two high-quality VLBA maps are available. To address the possibility of resolved-out flux in VLBA maps, we also estimate the synchrotron flux from the overall radio spectrum from a combination of different single-dish and interferometric surveys. With this method, we find typical magnetic field strengths of ${\sim}10$--$20\,\mathrm{mG}$ and minimum energies of ${\sim}5$--$10\,\mathrm{M}_{\odot}c^2$, with CSO-2s exhibiting a large range of minimum energies from $10^{-2}$--$10^{2}\,\mathrm{M_{\odot}c^2}$. Modeling of the X-ray inverse Compton emission by the same electron population generating the synchrotron emission shows that CSO-2s are likely within a factor of a few from minimum energy. 

A number of sources of systematic error that may significantly affect the minimum energy quantities of the CSO-2 population exist, including uncertainties in the component volume modeling (due to resolved-out flux or blended components), cutoffs of the injection electron distribution, volume filling factors, and baryonic contributions to the particle content. Furthermore, improvements to this analysis to reduce systematic errors in modeling require performing a full decomposition of components across $\geq3$ VLBI frequencies to improve determination of the individual component synchrotron spectrum. This would also allow us to place constraints on the existence of cooling breaks in each component and improve the synchrotron emission modeling.

If the cutoff in the CSO-2 size distribution at 500 pc is real, then the typical magnetic field strength of CSO-2s is able to explain their disappearance once the jet turns off. Deeper, lower frequency interferometric surveys should be able to identify an even larger sample of relic emission of once actively-accreting CSO-2s in the Local Universe. In addition, the identification of the smallest CSO-2s ($\leq 10$ pc) will provide strong constraints on theories of their origin and early-time evolution.

On the other hand, the typical CSO-2 minimum energies found by our analysis exceed those of \citetalias{2024ApJ...961..242R}. If CSOs are to be explained by the disruption of an evolved several solar mass star, then it must tap into a significant amount of the spin energy of the SMBH. In order to determine whether CSO-2s are truly triggered by discrete mass injections, it will be necessary to verify whether the size cutoff at ${\sim}500$ pc is real through deeper complete surveys. As more CSOs are identified \cite[e.g.,][]{Sheldahl_2025}, we plan to achieve a greater census of their components' magnetic field strengths and internal energy distributions.

%% Please use the acknowledgment and contribution environments. This will 
%% be anonomyized when the "anonymous" style option is used. 
\begin{acknowledgments}
We thank the anonymous referee for comments that helped clarify the presented methodology and encouraged a deeper discussion of the systematics present within the analyses. 

This work is supported by NSF grant AST2407603. TDS acknowledges support from the Caltech Doug and Betty Nickerson Fellowship. AGS acknowledges the support of the National Science Foundation Graduate Research Fellowship and a Giddings Fellowship at the Kavli Institute for Particle Astrophysics and Cosmology at Stanford. 

This paper depended on Very Long Baseline Array (VLBA) data obtained from the Radio Fundamental Catalog (DOI 10.25966/dhrk-zh08). The National Radio Astronomy Observatory is a facility of the National Science Foundation operated under cooperative agreement by Associated Universities, Inc. 

\end{acknowledgments}

\begin{contribution}
%%This section gives authors the space to recognize author contributions. The text inside this environment is NOT counted towards the total word quanta. At a minimum, manuscripts are expected to include this text:

T.D.S. was responsible for implementing the component fitting, spectral fitting, minimum energy calculations, and inverse Compton modeling and writing the manuscript. M.S.S.L.O. contributed to the derivations and methodological ideas for the analysis, and is responsible for collecting the data from the radio surveys. M.S.S.L.O., A.C.S.R., and A.G.S. provided useful discussions and checks on the analysis procedure and comments on the manuscript.

%% But authors are expected to provide more specific details, e.g. 
%%
%%SC was responsible for writing and submitting the manuscript.
%%WWM came up with the initial research concept and edited the manuscript.
%%OTS obtained the funding and edited the manuscript.
%%EBF provided the formal analysis and validation. He also edited the manuscript.
%%GEH Supervised the undergraduates, wrote the software and administers the project github and Zenodo repositories.
%%
%% Authors can use the Contributor Role Taxonomy (CRediT) at
%% for ideas on how write a good statement tailored to their needs.

\end{contribution}

\facilities{VLBA, VLA, ALMA, LOFAR, GMRT, MWA, WSRT, ASKAP, Planck, OVRO:40m, WISE, FLWO:2MASS, UKIRT, PS1, Sloan, XMM, CXO}

\software{Aegean-Tools \citep{2012MNRAS.422.1812H, 2018PASA...35...11H}, Astropy \citep{astropy:2013, astropy:2018, astropy:2022}, Xspec \citep{1999ascl.soft10005A}, AGNFitter \citep{2016ApJ...833...98C}, Imfit \citep{2015ApJ...799..226E}, Photutils \citep{2016ascl.soft09011B}, emcee \citep{2013PASP..125..306F}}


\clearpage
\appendix
\section{RFC VLBA Maps used for Individual Component Fitting}
Table \ref{tab:rfc_meta} shows the selected VLBA maps used from the Radio Fundamental Catalog necessary for decomposing the CSO-2s into individual components to compute volumes and computing component-wise spectral indices when a low and high frequency map are available.
\label{sec:rfcVLBAmaps}
\startlongtable
\begin{deluxetable*}{lcccl}
\tablecaption{
Frequencies and Observation Dates of the RFC VLBA Maps Used for Each Source.
%RFC VLBA map frequencies and observation dates used for each source.
%for which a high frequency is not listed
\label{tab:rfc_meta}}
\tablehead{
\colhead{J2000 Name} & \colhead{Low Freq. (GHz)} & \colhead{Low Date} & \colhead{High Freq. (GHz)} & \colhead{High Date}
}
\startdata
J0029+3456 & 4.9 & 1996-06-05 & 15.3\textasteriskcentered & 1996-07-10 \\
J0111+3906 & 4.9 & 1996-06-05 & 8.3\textasteriskcentered & 1996-06-07 \\
%J0131+5545 & 8.6 & 2010-08-29 & - & - \\
J0405+3803 & 2.3 & 2017-01-21 & - & - \\
J0713+4349 & 2.3 & 2018-06-03 & 8.7\textasteriskcentered & 2018-06-03 \\
J0741+2706 & 2.3 & 2014-06-09 & 4.8\textasteriskcentered & 2006-01-27 \\
%J0825+3919 & 4.1 & 2013-02-12 & - & - \\
J0855+5751 & 2.3 & 2018-05-07 & 4.8\textasteriskcentered & 2004-06-28 \\
J0943+1702 & 4.8 & 2006-01-28 & 8.7 & 2014-12-20 \\
J1035+5628 & 4.3 & 2018-05-22 & 7.6\textasteriskcentered & 2018-05-22 \\
J1111+1955 & 2.3 & 2018-03-26 & 8.7\textasteriskcentered & 2018-03-26 \\
J1120+1420 & 2.3 & 2018-04-08 & - & - \\
J1158+2450 & 8.3 & 1997-01-10 & 15.3\textasteriskcentered & 1999-05-21 \\
J1159+5820 & 4.8 & 2006-05-27 & - & - \\
J1227+3635 & 2.3 & 2014-05-31 & 4.8 & 2006-05-27 \\
J1244+4048 & 2.3 & 2018-07-01 & 8.7\textasteriskcentered & 2018-07-01 \\
J1247+6723 & 2.3 & 2018-07-01 & 8.7 & 2018-07-01 \\
J1313+5458 & 2.3 & 2006-12-18 & 4.8\textasteriskcentered & 2006-07-17 \\
J1326+3154 & 2.3 & 2018-08-10 & 4.8\textasteriskcentered & 2006-07-17 \\
%J1347+1217 & 4.9 & 1996-06-05 & - & - \\
J1400+6210 & 2.3 & 2017-06-15 & - & - \\
J1407+2827 & 8.4 & 2011-06-14 & 15.4\textasteriskcentered & 2011-12-12 \\
J1414+4554 & 2.3 & 2018-09-20 & 8.7\textasteriskcentered & 2018-09-20 \\
J1434+4236 & 4.8 & 2006-08-03 & - & - \\
J1440+6108 & 4.8 & 2006-08-03 & - & - \\
J1511+0518 & 8.4 & 2012-02-25 & 15.4\textasteriskcentered & 2012-12-23 \\
J1609+2641 & 2.3 & 2017-08-01 & 8.6\textasteriskcentered & 2017-08-01 \\
J1644+2536 & 2.3 & 2015-01-23 & 4.8 & 2006-04-03 \\
J1734+0926 & 2.3 & 2018-09-24 & 8.7\textasteriskcentered & 2018-09-24 \\
J1735+5049 & 8.7 & 2018-08-10 & - & - \\
J1816+3457 & 4.3 & 2018-09-07 & 7.6\textasteriskcentered & 2018-09-07 \\
J1915+6548 & 2.3 & 2018-04-08 & 8.7\textasteriskcentered & 2018-04-08 \\
J1944+5448 & 2.3 & 2014-05-31 & 8.7\textasteriskcentered & 2014-05-31 \\
J1945+7055 & 2.3 & 2018-11-03 & 8.7\textasteriskcentered & 2018-11-03 \\
J2022+6136 & 7.6 & 2015-06-12 & 15.4 & 2013-07-22 \\
J2203+1007 & 2.3 & 2017-01-16 & - & - \\
J2355+4950 & 2.3 & 2017-04-25 & - & - \\
\enddata
\end{deluxetable*}
\tablecomments{For sources without a high-frequency entry, we only used a single map.
High frequencies with an asterisk are those whose maps were convolved to match the resolution of the low-frequency map, for improved component matching.}
\section{Radio Surveys used for Spectral Fits}
Table \ref{tab:surveys} lists the primary catalogues from which the broadband synchrotron spectrum was constructed for each CSO-2, which was used to perform double power law fits in the spectral analysis.

\startlongtable
\begin{deluxetable*}{c|c|c|c}
\tablehead{
\colhead{Catalogue} &
\colhead{Frequency ($\mathrm{MHz}$)} &
\colhead{PSF FWHM ($''$)} &
%\colhead{Noise SD ($\mathrm{Jy\ PSF^{-1}}$)} &
\colhead{Reference}
}
\tablecaption{%
Selected Properties of the Surveys through which We Assembled Radio Spectra.
%the first range is for the lowest frequency, while the second range is for the highest frequency.}
%we first list the range
%list the ranges for the lower and upper frequencies and from the most compact to the most extended configurations.
%The image noise standard deviation (SD) unit, $\mathrm{Jy\ PSF^{-1}}$, is also known as `$\mathrm{Jy\ beam^{-1}}$'.
\label{tab:surveys}
}
\startdata
LoLSS & 54 & 15 &  \textcite{deGasperin12023} \\
VLSSr & 74 & 75 &  \textcite{Lane12012} \\
LoTSS & 144 & 6 &  \textcite{Shimwell12022} \\
TGSS ADR & 150 & 25 & \textcite{Intema12017} \\
GLEAM & 200 & 120 & \textcite{HurleyWalker12017} \\
WENSS & 325 & 54 & \textcite{Rengelink11997} \\
VCSS & 340 & 15 &  \textcite{Polisensky12016} \\
RACS-low & 887.5 & 15 & \textcite{Hale12021} \\
RACS-mid & 1367.5 & 10 & \textcite{Duchesne12024} \\
NVSS & 1400 & 45 & \textcite{Condon11998} \\
FIRST & 1400 & 5 &  \textcite{Becker11995} \\
RACS-high & 1655.5 & 8 & \textcite{Duchesne12025} \\
VLASS & 3000 & 2.5 &  \textcite{Gordon12021} \\
%AT 5 GHz & ? & ? & ? & \\
%AT 8 GHz & ? & ? & ? & \\
%AT 20 GHz & ? & ? & ? & \\
CRATES & 4800--8400 & 0.4--0.2 & \textcite{2007ApJS..171...61H} \\
PCNT & 28,400 & 1945 &  \textcite{Planck12018} \\
VLA CC & 330--43,200 & 5.6--0.04 & \\
ACC & 100,000--442,500 & (12--0.041)--(2.6--0.0091) & \textcite{Bonato12019} \\
\enddata
\tablecomments{We quantify angular resolution through the full width at half maximum of the point spread function.
LoTSS imagery is also available at a resolution of $20''$.
For the CRATES and VLA CC surveys, we list the A-configuration PSF FWHMs.
For the ACC, in parentheses, we list FWHMs for the most compact to the most extended configurations; the two ranges are for the lowest and highest frequencies, respectively.}
\end{deluxetable*}

\section{Synchrotron Minimum Energy Derivation}
\label{sec:deriv}
%We assume that the underlying particle distribution
In this appendix, we derive the `minimum energy' expressions used in this paper's CSO component modeling. 
We assume that the component lepton (i.e. electron and positron) population generating the synchrotron photons follows a truncated power law distribution in their Lorentz factors (or, equivalently, in their energies): $f_{\Gamma}(\gamma) = C_e\gamma^{p}$, where $p < 0$ and $\gamma_{\min}\leq \gamma\leq\gamma_{\max}$.
It is assumed that $p$ is related to the synchrotron spectral index $\alpha$ by $p=2\alpha-1$.
Here, $C_e$ is a normalization constant that ensures that $f_\Gamma(\gamma)$ integrates to 1 over the full range of Lorentz factors.
In this paper, we have assumed $\gamma_{\min}=10$ and $\gamma_{\max}=10^{5}.$
%is comprised in
The total energy density comprises contributions from particles (electrons and positrons on the one hand, and protons on the other) and from magnetic fields:
\begin{equation}
u = u_e + u_p + u_B = (1+K)u_e + u_B.
\end{equation}
Here, the protons contribute an energy density $K u_e$; the magnetic field energy density $u_B = B^2/2\mu_0$.
In this paper, we take the jets to be leptonic and ignore entrainment, so that the proton contribution to the total energy is negligible ($K \approx 0$).
%We take the jets to be leptonic in this paper, in the sense that the proton contribution to the total energy is negligible ($K=0$).
We can express the lepton energy density as the total (i.e. rest plus kinetic) energy of the $N_e$ leptons, which have some mean Lorentz factor $\mathbb{E}[\Gamma] \coloneqq \int \gamma f_\Gamma(\gamma)\ \mathrm{d}\gamma$, divided by the product of the plasma's volume filling fraction $\eta$ and the lobe volume $V$:
%taken with respect to $f_{\Gamma}(\gamma)$; uppercase is used to denote the random variable
\begin{equation}
u_e = \frac{N_em_ec^2\mathbb{E}[\Gamma]}{\eta V}.
\end{equation}
In this paper, we assume $\eta = 1$.
%where $\eta$ is the volume filling factor, nominally assumed to be $1$ in this paper.
Neglecting synchrotron radiation from protons, the total number of electrons and positrons, $N_e$, is given by the ratio of the total synchrotron luminosity to the pitch angle--averaged mean synchrotron power per lepton:
%$L_\mathrm{s}$, $\mathbb{E}[P_\mathrm{s}]$
\begin{equation}
N_e = \frac{L_\mathrm{s}}{\mathbb{E}[P_\mathrm{s}]} = \frac{3 \mu_0 L_\mathrm{s}}{2 \mathbb{E}[\Gamma^2-1] \sigma_\mathrm{T} c B^2}.
\end{equation}
We compute $L_\mathrm{s}$ by assuming that the lobe produces, at least internally, a synchrotron spectrum with a power law profile; the exponent is the optically thin spectral index $\alpha$:
%the source intrinsically follows a power law profile set by the optically thin spectral index:
\begin{equation}
L_{\nu}(\nu) = L_0\left(\frac{\nu}{\nu_0}\right)^{\alpha}.
\label{eq:lum_prof}
\end{equation}
Here, $L_0$ is the internally generated (rather than outwardly radiated) luminosity density at reference rest-frame frequency $\nu_0$.
%Here, $L_0$ and $\nu_0$ are the redshift-corrected reference luminosity density and frequency well into the optically thin regime.
We integrate $L_\nu(\nu)$ between the characteristic frequencies (Equation \ref{eq:char_freq}) of the minimum and maximum leptonic Lorentz factors $\gamma_{\min}$ and $\gamma_{\max}$.
For $\alpha\neq-1$, we obtain
%For simplicity, we show the result for $\alpha\neq-1$:
\begin{equation}
L_\mathrm{s} = C_1B^{\alpha+1};\, C_1\equiv\frac{L_0\nu_0}{\alpha+1}\left(\frac{3e}{16m_e\nu_0}\right)^{\alpha+1}\left(\gamma_{\max}^{2\alpha+2}-\gamma_{\min}^{2\alpha+2}\right).
\end{equation}
We rewrite the total energy density as
\begin{equation}
u(B) = \frac{C_1C_2C_3 B^{\alpha+1}}{B^2} + \frac{B^2}{2\mu_0},
\label{eq:min_energy}
\end{equation}
with $C_2$ being the lobe's rest-frame particulate energy density per lepton, and $C_3$ encompassing all other factors:
\begin{equation}
C_2 \equiv \frac{m_ec^2\mathbb{E}[\Gamma](1+K)}{\eta V};\, C_3 \equiv \frac{3\mu_0}{2\mathbb{E}[\Gamma^2-1]\sigma_\mathrm{T}c}.
\end{equation}
One obtains the minimum energy magnetic field strength by solving $\frac{\mathrm{d}u}{\mathrm{d}B}(B_\mathrm{ME}) = 0$.
The result is
%differentiating:
\begin{equation}
B_\mathrm{ME} = ((1-\alpha)\mu_0C_1C_2C_3)^{\frac{1}{3-\alpha}},
\label{eq:b_min}
\end{equation}
for $\alpha\neq -1$. The minimum energy in a synchrotron-emitting region can then be found through
%by taking
\begin{equation}
E_{\min} = u(B_\mathrm{ME})\eta V.
\label{eq:e_min}
\end{equation}

\section{CSO-2 Fits, Minimum Energies, and Magnetic Fields}

Table \ref{tab:ind_results} lists the individual components fitted to each CSO-2 from which volumes were computed. Minimum energies and magnetic field strengths obtained from the individual component analysis are also provided for CSO-2s where two VLBA maps were available. Table \ref{tab:spec_results} lists the total minimum energy and average magnetic field strengths derived from the spectral analysis, along with relevant information describing the shape and normalization of the optically thin part of the broadband radio spectrum.

\startlongtable
\begin{deluxetable*}{lccccccc}
\tablecaption{Fits of Individual Components of CSO-2s} \label{tab:ind_results}
\tablehead{
\colhead{Name} & \colhead{Component} & \colhead{$a$ (mas)} & \colhead{$b$ (mas)} & \colhead{$F_\nu$ (mJy)} & \colhead{$\alpha$} & \colhead{$B_{\mathrm{ME}}$ (mG)} & \colhead{$E_{\min}$ ($M_{\odot}c^2$)} \\
}
\startdata
J0029+3456 & C1 & $3.236^{+0.014}_{-0.014}$ & $1.19^{+0.02}_{-0.02}$ & $788.1^{+1.7}_{-1.8}$ & $-0.680^{+0.006}_{-0.007}$ & $19.40^{+0.16}_{-0.16}$ & $5.07^{+0.07}_{-0.07}$ \\
 & C2 & $2.79^{+0.05}_{-0.05}$ & $1.409^{+0.019}_{-0.019}$ & $274.3^{+2.0}_{-1.7}$ & $-1.49^{+0.04}_{-0.03}$ & $41.3^{+1.9}_{-1.9}$ & $18.3^{+1.6}_{-1.6}$ \\
 & C3 & $1.71^{+0.11}_{-0.08}$ & $1.59^{+0.05}_{-0.05}$ & $133.1^{+1.6}_{-1.8}$ & $-1.08^{+0.05}_{-0.05}$ & $23.5^{+1.9}_{-1.7}$ & $3.5^{+0.6}_{-0.5}$ \\
J0111+3906 & C1 & $1.580^{+0.008}_{-0.008}$ & $0.16^{+0.11}_{-0.09}$ & $869^{+2}_{-2}$ & $-0.723^{+0.006}_{-0.006}$ & $59^{+17}_{-8}$ & $1.7^{+0.5}_{-0.6}$ \\
 & C2 & $1.02^{+0.03}_{-0.02}$ & $0.51^{+0.09}_{-0.11}$ & $511^{+2}_{-2}$ & $-0.991^{+0.011}_{-0.011}$ & $57^{+5}_{-3}$ & $2.6^{+0.3}_{-0.4}$ \\
J0405+3803 & C1 & $6.67^{+0.03}_{-0.03}$ & $2.74^{+0.05}_{-0.04}$ & $446.1^{+1.4}_{-1.2}$ & - & - & - \\
 & C2 & $3.90^{+0.09}_{-0.10}$ & $2.56^{+0.07}_{-0.07}$ & $130.2^{+1.3}_{-1.3}$ & - & - & - \\
 & C3 & $2.2^{+0.3}_{-0.3}$ & $1.3^{+0.5}_{-0.7}$ & $37.5^{+1.0}_{-0.9}$ & - & - & - \\
 & C4 & $9.8^{+0.4}_{-0.4}$ & $3.04^{+0.17}_{-0.17}$ & $58^{+2}_{-2}$ & - & - & - \\
J0713+4349 & C1 & $2.09^{+0.05}_{-0.05}$ & $1.17^{+0.05}_{-0.05}$ & $230.3^{+1.1}_{-1.0}$ & $-0.98^{+0.03}_{-0.03}$ & $20.0^{+0.8}_{-0.9}$ & $2.30^{+0.17}_{-0.18}$ \\
 & C2 & $2.316^{+0.008}_{-0.008}$ & $1.713^{+0.016}_{-0.014}$ & $1107.7^{+1.0}_{-1.1}$ & $-0.666^{+0.006}_{-0.005}$ & $18.62^{+0.09}_{-0.09}$ & $4.40^{+0.03}_{-0.03}$ \\
 & C3 & $2.72^{+0.02}_{-0.02}$ & $0.50^{+0.04}_{-0.04}$ & $608.2^{+1.0}_{-1.1}$ & $-0.504^{+0.008}_{-0.009}$ & $22.8^{+0.7}_{-0.6}$ & $1.90^{+0.07}_{-0.09}$ \\
J0741+2706 & C1 & $2.622^{+0.005}_{-0.005}$ & $1.826^{+0.014}_{-0.012}$ & $856.7^{+0.4}_{-0.4}$ & $-0.574^{+0.002}_{-0.002}$ & $17.45^{+0.05}_{-0.06}$ & $8.97^{+0.05}_{-0.04}$ \\
 & C2 & $8.37^{+0.10}_{-0.10}$ & $5.42^{+0.11}_{-0.11}$ & $68.4^{+0.6}_{-0.5}$ & $-1.01^{+0.05}_{-0.05}$ & $5.7^{+0.5}_{-0.5}$ & $25^{+4}_{-4}$ \\
J0855+5751 & C1 & $8.18^{+0.04}_{-0.04}$ & $5.32^{+0.03}_{-0.03}$ & $200.2^{+0.7}_{-0.8}$ & $-1.50^{+0.02}_{-0.02}$ & $16.0^{+0.5}_{-0.5}$ & $0.059^{+0.003}_{-0.003}$ \\
 & C2 & $8.29^{+0.04}_{-0.04}$ & $4.52^{+0.03}_{-0.03}$ & $186.1^{+0.7}_{-0.8}$ & $-1.67^{+0.03}_{-0.03}$ & $20.2^{+0.7}_{-0.7}$ & $0.075^{+0.005}_{-0.005}$ \\
J0943+1702 & C1 & $1.23^{+0.04}_{-0.04}$ & $0.07^{+0.06}_{-0.04}$ & $193.2^{+0.9}_{-0.9}$ & $-0.163^{+0.009}_{-0.008}$ & - & - \\
 & C2 & $2.57^{+0.07}_{-0.08}$ & $1.11^{+0.07}_{-0.09}$ & $71.6^{+1.0}_{-1.1}$ & $-1.32^{+0.03}_{-0.03}$ & $43^{+2}_{-2}$ & $32^{+4}_{-4}$ \\
 & C3 & $0.7^{+0.2}_{-0.3}$ & $0.2^{+0.2}_{-0.2}$ & $27.4^{+0.9}_{-0.9}$ & $-1.13^{+0.06}_{-0.06}$ & $70^{+40}_{-20}$ & $1.4^{+1.1}_{-0.9}$ \\
J1035+5628 & C1 & $2.132^{+0.008}_{-0.009}$ & $1.14^{+0.02}_{-0.02}$ & $644.3^{+0.9}_{-0.9}$ & $-0.924^{+0.004}_{-0.004}$ & $27.7^{+0.3}_{-0.3}$ & $3.70^{+0.06}_{-0.07}$ \\
 & C2 & $2.905^{+0.015}_{-0.016}$ & $1.49^{+0.03}_{-0.03}$ & $451.4^{+1.0}_{-1.0}$ & $-0.753^{+0.005}_{-0.006}$ & $16.57^{+0.16}_{-0.15}$ & $3.33^{+0.05}_{-0.05}$ \\
J1111+1955 & C1 & $2.775^{+0.007}_{-0.008}$ & $1.38^{+0.05}_{-0.05}$ & $625.8^{+0.6}_{-0.6}$ & $-0.895^{+0.002}_{-0.003}$ & $18.2^{+0.2}_{-0.2}$ & $1.44^{+0.03}_{-0.03}$ \\
 & C2 & $3.236^{+0.014}_{-0.014}$ & $1.21^{+0.04}_{-0.05}$ & $485.0^{+0.5}_{-0.6}$ & $-1.037^{+0.003}_{-0.004}$ & $19.6^{+0.3}_{-0.2}$ & $1.76^{+0.05}_{-0.05}$ \\
J1120+1420 & C1 & $5.82^{+0.15}_{-0.15}$ & $3.38^{+0.07}_{-0.08}$ & $347^{+4}_{-4}$ & - & - & - \\
 & C2 & $5.8^{+0.3}_{-0.3}$ & $2.1^{+0.4}_{-0.6}$ & $170^{+5}_{-5}$ & - & - & - \\
 & C3 & $7.2^{+0.2}_{-0.2}$ & $6.7^{+0.2}_{-0.2}$ & $262^{+5}_{-5}$ & - & - & - \\
 & C4 & $5.26^{+0.15}_{-0.15}$ & $0.6^{+0.5}_{-0.4}$ & $192^{+3}_{-2}$ & - & - & - \\
 & C5 & $10.72^{+0.18}_{-0.14}$ & $7.99^{+0.20}_{-0.19}$ & $430^{+5}_{-6}$ & - & - & - \\
J1158+2450 & C1 & $0.62^{+0.03}_{-0.03}$ & $0.57^{+0.03}_{-0.04}$ & $143.5^{+1.7}_{-1.9}$ & $-0.69^{+0.03}_{-0.03}$ & $34.9^{+1.7}_{-1.5}$ & $0.062^{+0.005}_{-0.005}$ \\
 & C2 & $0.4^{+0.7}_{-0.3}$ & $0.09^{+0.14}_{-0.07}$ & $16.1^{+1.3}_{-1.3}$ & $0.78^{+0.14}_{-0.13}$ & - & - \\
 & C3 & $2.29^{+0.11}_{-0.09}$ & $2.02^{+0.09}_{-0.12}$ & $69^{+2}_{-2}$ & $-1.02^{+0.09}_{-0.08}$ & $16^{+2}_{-2}$ & $0.56^{+0.15}_{-0.12}$ \\
 & C4 & $2.7^{+0.3}_{-0.3}$ & $1.63^{+0.16}_{-0.17}$ & $40^{+3}_{-3}$ & $-1.8^{+0.2}_{-0.2}$ & $46^{+15}_{-11}$ & $4^{+3}_{-2}$ \\
J1159+5820$^{\star}$ & C1 & $2.94^{+0.05}_{-0.04}$ & $2.42^{+0.04}_{-0.04}$ & $130.4^{+1.1}_{-1.1}$ & - & - & - \\
 & C2 & $4.71^{+0.07}_{-0.07}$ & $2.69^{+0.05}_{-0.05}$ & $141.3^{+1.5}_{-1.4}$ & - & - & - \\
J1227+3635$^{\star}$ & C1 & $3.315^{+0.012}_{-0.011}$ & $2.269^{+0.013}_{-0.013}$ & $879.4^{+0.9}_{-0.8}$ & $-0.979^{+0.004}_{-0.003}$ & $33.56^{+0.16}_{-0.18}$ & $82.7^{+0.8}_{-0.8}$ \\
 & C2 & $3.31^{+0.06}_{-0.06}$ & $1.47^{+0.10}_{-0.09}$ & $190.3^{+1.0}_{-1.0}$ & $-1.18^{+0.02}_{-0.03}$ & $34.4^{+1.7}_{-1.2}$ & $46^{+4}_{-3}$ \\
 & C3 & $5.68^{+0.11}_{-0.11}$ & $3.04^{+0.10}_{-0.12}$ & $102.0^{+1.1}_{-1.1}$ & $-2.05^{+0.08}_{-0.08}$ & $58^{+6}_{-5}$ & $720^{+140}_{-110}$ \\
 & C4 & $3.40^{+0.09}_{-0.09}$ & $1.0^{+0.3}_{-0.3}$ & $103.9^{+1.1}_{-1.0}$ & $-1.11^{+0.03}_{-0.04}$ & $31^{+4}_{-3}$ & $23^{+5}_{-5}$ \\
 & C5 & $2.35^{+0.10}_{-0.11}$ & $0.6^{+0.2}_{-0.2}$ & $93.2^{+0.9}_{-1.0}$ & $-1.12^{+0.04}_{-0.04}$ & $40^{+5}_{-4}$ & $12^{+3}_{-3}$ \\
J1244+4048$^{\star}$ & C1 & $4.85^{+0.09}_{-0.10}$ & $1.92^{+0.10}_{-0.12}$ & $153.9^{+1.8}_{-1.6}$ & $-0.16^{+0.04}_{-0.04}$ & - & - \\
 & C2 & $5.71^{+0.08}_{-0.08}$ & $0.13^{+0.08}_{-0.05}$ & $194.9^{+1.6}_{-1.6}$ & $-0.81^{+0.18}_{-0.13}$ & $21^{+4}_{-3}$ & $2.6^{+1.0}_{-0.6}$ \\
 & C3 & $8.53^{+0.08}_{-0.07}$ & $5.21^{+0.05}_{-0.05}$ & $303^{+2}_{-2}$ & $-1.24^{+0.11}_{-0.15}$ & $12^{+3}_{-2}$ & $100^{+50}_{-30}$ \\
 & C4 & $4.87^{+0.06}_{-0.06}$ & $3.45^{+0.05}_{-0.05}$ & $242.6^{+1.4}_{-1.4}$ & $-1.15^{+0.10}_{-0.13}$ & $14^{+3}_{-2}$ & $34^{+13}_{-8}$ \\
J1247+6723 & C1 & $1.39^{+0.05}_{-0.05}$ & $0.09^{+0.07}_{-0.06}$ & $146.4^{+0.4}_{-0.4}$ & $-0.516^{+0.003}_{-0.003}$ & $37^{+12}_{-6}$ & $0.007^{+0.002}_{-0.002}$ \\
 & C2 & $1.85^{+0.05}_{-0.05}$ & $0.07^{+0.06}_{-0.04}$ & $128.5^{+0.4}_{-0.4}$ & $-0.643^{+0.004}_{-0.004}$ & $32^{+7}_{-5}$ & $0.007^{+0.002}_{-0.002}$ \\
J1313+5458 & C1 & $5.64^{+0.06}_{-0.06}$ & $4.47^{+0.03}_{-0.04}$ & $379.3^{+1.8}_{-1.7}$ & $-1.34^{+0.03}_{-0.02}$ & $16.3^{+0.5}_{-0.5}$ & $58^{+4}_{-4}$ \\
 & C2 & $3.67^{+0.03}_{-0.03}$ & $2.72^{+0.05}_{-0.05}$ & $498.7^{+1.4}_{-1.4}$ & $-0.759^{+0.010}_{-0.010}$ & $11.61^{+0.15}_{-0.15}$ & $8.45^{+0.18}_{-0.18}$ \\
J1326+3154 & C1 & $4.80^{+0.07}_{-0.05}$ & $4.66^{+0.06}_{-0.08}$ & $689^{+5}_{-5}$ & $-0.502^{+0.013}_{-0.012}$ & $7.44^{+0.08}_{-0.07}$ & $5.50^{+0.10}_{-0.10}$ \\
 & C2 & $6.36^{+0.06}_{-0.07}$ & $5.60^{+0.06}_{-0.06}$ & $1027^{+6}_{-6}$ & $-0.484^{+0.010}_{-0.009}$ & $6.88^{+0.05}_{-0.05}$ & $9.55^{+0.15}_{-0.14}$ \\
J1400+6210$^{\star}$ & C1 & $6.87^{+0.04}_{-0.04}$ & $3.40^{+0.04}_{-0.03}$ & $922^{+4}_{-4}$ & - & - & - \\
 & C2 & $5.63^{+0.07}_{-0.08}$ & $1.31^{+0.12}_{-0.12}$ & $363^{+3}_{-4}$ & - & - & - \\
 & C3 & $7.97^{+0.18}_{-0.18}$ & $2.52^{+0.14}_{-0.12}$ & $276^{+4}_{-5}$ & - & - & - \\
 & C4 & $4.18^{+0.04}_{-0.04}$ & $2.78^{+0.04}_{-0.04}$ & $615^{+3}_{-3}$ & - & - & - \\
 & C5 & $10.4^{+0.6}_{-0.5}$ & $6.6^{+0.3}_{-0.3}$ & $127^{+5}_{-5}$ & - & - & - \\
J1407+2827 & C1 & $0.966^{+0.020}_{-0.020}$ & $0.76^{+0.05}_{-0.06}$ & $81.5^{+0.9}_{-0.8}$ & $-1.74^{+0.05}_{-0.04}$ & $88^{+5}_{-5}$ & $0.078^{+0.009}_{-0.009}$ \\
 & C2 & $0.77^{+0.06}_{-0.06}$ & $0.47^{+0.05}_{-0.06}$ & $56.5^{+0.7}_{-0.7}$ & $-2.45^{+0.08}_{-0.09}$ & $203^{+19}_{-16}$ & $0.13^{+0.03}_{-0.03}$ \\
 & C3 & $1.491^{+0.003}_{-0.002}$ & $0.8294^{+0.0011}_{-0.0012}$ & $1225.7^{+0.7}_{-0.7}$ & $-1.848^{+0.002}_{-0.003}$ & $146.2^{+0.5}_{-0.4}$ & $0.491^{+0.003}_{-0.003}$ \\
J1414+4554 & C1 & $5.41^{+0.02}_{-0.02}$ & $2.39^{+0.03}_{-0.03}$ & $192.4^{+0.5}_{-0.4}$ & $-0.923^{+0.005}_{-0.005}$ & $8.72^{+0.07}_{-0.07}$ & $0.712^{+0.011}_{-0.011}$ \\
 & C2 & $4.19^{+0.03}_{-0.04}$ & $1.89^{+0.03}_{-0.03}$ & $130.6^{+0.4}_{-0.4}$ & $-0.835^{+0.008}_{-0.008}$ & $8.57^{+0.10}_{-0.10}$ & $0.336^{+0.007}_{-0.007}$ \\
J1434+4236 & C1 & $2.950^{+0.017}_{-0.018}$ & $1.82^{+0.02}_{-0.02}$ & $210.6^{+0.7}_{-0.7}$ & - & - & - \\
 & C2 & $10.4^{+1.0}_{-0.9}$ & $5.8^{+0.5}_{-0.4}$ & $28^{+2}_{-2}$ & - & - & - \\
J1440+6108 & C1 & $2.05^{+0.10}_{-0.09}$ & $1.41^{+0.12}_{-0.15}$ & $37.4^{+0.8}_{-0.7}$ & - & - & - \\
 & C2 & $3.30^{+0.07}_{-0.06}$ & $2.78^{+0.09}_{-0.08}$ & $59.0^{+0.8}_{-0.9}$ & - & - & - \\
J1511+0518 & C1 & $1.264^{+0.003}_{-0.003}$ & $0.559^{+0.006}_{-0.006}$ & $541.6^{+0.4}_{-0.5}$ & $-0.495^{+0.002}_{-0.002}$ & $39.9^{+0.2}_{-0.2}$ & $0.0286^{+0.0002}_{-0.0002}$ \\
 & C2 & $0.85^{+0.03}_{-0.03}$ & $0.456^{+0.015}_{-0.015}$ & $95.0^{+0.4}_{-0.4}$ & $-1.30^{+0.02}_{-0.02}$ & $66^{+2}_{-2}$ & $0.0242^{+0.0016}_{-0.0014}$ \\
 & C3 & $1.59^{+0.09}_{-0.07}$ & $0.12^{+0.09}_{-0.07}$ & $25.8^{+0.4}_{-0.4}$ & $0.52^{+0.03}_{-0.03}$ & - & - \\
J1609+2641 & C1 & $6.75^{+0.02}_{-0.02}$ & $3.378^{+0.014}_{-0.015}$ & $1353^{+3}_{-3}$ & $-1.348^{+0.005}_{-0.005}$ & $21.15^{+0.14}_{-0.14}$ & $59.6^{+0.8}_{-0.8}$ \\
 & C2 & $4.233^{+0.012}_{-0.012}$ & $2.853^{+0.010}_{-0.011}$ & $1710^{+2}_{-2}$ & $-0.959^{+0.002}_{-0.002}$ & $17.50^{+0.06}_{-0.06}$ & $16.52^{+0.10}_{-0.10}$ \\
J1644+2536 & C1 & $3.61^{+0.05}_{-0.05}$ & $2.32^{+0.02}_{-0.03}$ & $191.1^{+0.6}_{-0.7}$ & $-1.37^{+0.02}_{-0.03}$ & $20.7^{+0.7}_{-0.6}$ & $17.1^{+1.1}_{-0.9}$ \\
 & C2 & $3.58^{+0.11}_{-0.10}$ & $0.10^{+0.07}_{-0.04}$ & $87.1^{+0.7}_{-0.7}$ & $1.238^{+0.011}_{-0.012}$ & - & - \\
 & C3 & $8.34^{+0.11}_{-0.11}$ & $1.08^{+0.10}_{-0.11}$ & $113.3^{+0.9}_{-0.9}$ & $-0.79^{+0.03}_{-0.03}$ & $7.5^{+0.4}_{-0.3}$ & $4.3^{+0.4}_{-0.4}$ \\
J1734+0926 & C1 & $2.18^{+0.04}_{-0.04}$ & $1.19^{+0.04}_{-0.05}$ & $606.9^{+1.0}_{-1.1}$ & $-0.804^{+0.003}_{-0.003}$ & $22.7^{+0.5}_{-0.4}$ & $5.5^{+0.2}_{-0.2}$ \\
 & C2 & $1.75^{+0.05}_{-0.05}$ & $1.393^{+0.017}_{-0.018}$ & $749.9^{+1.1}_{-1.1}$ & $-0.717^{+0.002}_{-0.002}$ & $23.1^{+0.3}_{-0.3}$ & $5.11^{+0.12}_{-0.12}$ \\
J1735+5049 & C1 & $0.426^{+0.004}_{-0.004}$ & $0.151^{+0.008}_{-0.009}$ & $499.4^{+0.5}_{-0.5}$ & - & - & - \\
 & C2 & $0.854^{+0.009}_{-0.010}$ & $0.497^{+0.009}_{-0.009}$ & $183.8^{+0.6}_{-0.6}$ & - & - & - \\
J1816+3457 & C1 & $4.918^{+0.019}_{-0.018}$ & $2.979^{+0.011}_{-0.012}$ & $214.7^{+0.4}_{-0.4}$ & $-1.193^{+0.008}_{-0.008}$ & $14.79^{+0.18}_{-0.17}$ & $4.18^{+0.09}_{-0.09}$ \\
 & C2 & $3.95^{+0.06}_{-0.06}$ & $2.68^{+0.03}_{-0.03}$ & $63.5^{+0.4}_{-0.4}$ & $-0.94^{+0.02}_{-0.02}$ & $8.5^{+0.3}_{-0.3}$ & $0.89^{+0.06}_{-0.06}$ \\
J1915+6548 & C1 & $1.87^{+0.16}_{-0.17}$ & $1.2^{+0.3}_{-0.4}$ & $52.3^{+0.7}_{-0.7}$ & $-1.36^{+0.11}_{-0.11}$ & $23^{+5}_{-3}$ & $2.0^{+0.8}_{-0.6}$ \\
 & C2 & $2.11^{+0.02}_{-0.02}$ & $1.873^{+0.015}_{-0.018}$ & $473.3^{+0.7}_{-0.8}$ & $-0.791^{+0.004}_{-0.004}$ & $16.17^{+0.12}_{-0.11}$ & $2.84^{+0.03}_{-0.03}$ \\
J1944+5448 & C1 & $4.400^{+0.007}_{-0.006}$ & $2.227^{+0.009}_{-0.007}$ & $778.4^{+0.6}_{-0.6}$ & $-0.593^{+0.005}_{-0.005}$ & $10.509^{+0.018}_{-0.020}$ & $1.629^{+0.004}_{-0.004}$ \\
 & C2 & $3.921^{+0.015}_{-0.015}$ & $1.35^{+0.02}_{-0.02}$ & $324.1^{+0.5}_{-0.5}$ & $-0.732^{+0.014}_{-0.016}$ & $11.27^{+0.17}_{-0.15}$ & $0.767^{+0.018}_{-0.017}$ \\
J1945+7055 & C1 & $3.06^{+0.02}_{-0.02}$ & $1.66^{+0.02}_{-0.02}$ & $426.4^{+1.0}_{-1.1}$ & $-0.843^{+0.009}_{-0.011}$ & $14.97^{+0.19}_{-0.18}$ & $0.109^{+0.002}_{-0.002}$ \\
 & C2 & $3.38^{+0.06}_{-0.07}$ & $0.90^{+0.14}_{-0.17}$ & $136.3^{+1.1}_{-1.1}$ & $-0.67^{+0.02}_{-0.03}$ & $11.3^{+0.8}_{-0.6}$ & $0.035^{+0.003}_{-0.004}$ \\
 & C3 & $5.09^{+0.09}_{-0.09}$ & $1.44^{+0.12}_{-0.11}$ & $99.2^{+1.0}_{-1.0}$ & $-0.165^{+0.016}_{-0.015}$ & - & - \\
 & C4 & $1.95^{+0.12}_{-0.14}$ & $1.1^{+0.4}_{-0.4}$ & $57.2^{+0.8}_{-0.8}$ & $-0.71^{+0.05}_{-0.05}$ & $11^{+2}_{-1}$ & $0.016^{+0.004}_{-0.004}$ \\
 & C5 & $4.22^{+0.12}_{-0.11}$ & $2.47^{+0.10}_{-0.10}$ & $72.8^{+0.9}_{-0.9}$ & $-1.24^{+0.09}_{-0.09}$ & $12.0^{+1.5}_{-1.4}$ & $0.19^{+0.05}_{-0.04}$ \\
J2022+6136 & C1 & $0.7476^{+0.0012}_{-0.0013}$ & $0.308^{+0.017}_{-0.017}$ & $1803.3^{+1.0}_{-0.9}$ & $-0.2091^{+0.0008}_{-0.0008}$ & $137^{+3}_{-3}$ & $0.88^{+0.02}_{-0.03}$ \\
 & C2 & $0.703^{+0.007}_{-0.005}$ & $0.57^{+0.03}_{-0.03}$ & $630.1^{+1.0}_{-1.1}$ & $-1.287^{+0.003}_{-0.003}$ & $91.2^{+1.6}_{-1.5}$ & $0.58^{+0.02}_{-0.02}$ \\
 & C3 & $0.795^{+0.007}_{-0.007}$ & $0.17^{+0.10}_{-0.10}$ & $287.5^{+1.1}_{-0.9}$ & $-0.253^{+0.005}_{-0.006}$ & - & - \\
 & C4 & $1.10^{+0.15}_{-0.12}$ & $0.3^{+0.3}_{-0.2}$ & $32.5^{+1.1}_{-1.1}$ & $-1.81^{+0.08}_{-0.08}$ & $90^{+30}_{-20}$ & $0.5^{+0.3}_{-0.3}$ \\
J2203+1007 & C1 & $3.11^{+0.04}_{-0.04}$ & $1.65^{+0.17}_{-0.18}$ & $121.5^{+0.5}_{-0.5}$ & - & - & - \\
 & C2 & $1.59^{+0.11}_{-0.12}$ & $0.3^{+0.3}_{-0.2}$ & $80.0^{+0.4}_{-0.4}$ & - & - & - \\
J2355+4950 & C1 & $3.22^{+0.06}_{-0.07}$ & $1.84^{+0.07}_{-0.07}$ & $154.8^{+1.7}_{-1.6}$ & - & - & - \\
 & C2 & $4.96^{+0.16}_{-0.14}$ & $3.97^{+0.14}_{-0.14}$ & $111^{+2}_{-2}$ & - & - & - \\
 & C3 & $5.38^{+0.17}_{-0.19}$ & $4.72^{+0.17}_{-0.16}$ & $112^{+2}_{-2}$ & - & - & - \\
 & C4 & $2.725^{+0.010}_{-0.008}$ & $1.386^{+0.013}_{-0.013}$ & $1187.0^{+1.4}_{-1.4}$ & - & - & - \\
 & C5 & $2.11^{+0.07}_{-0.07}$ & $2.00^{+0.07}_{-0.09}$ & $134.2^{+1.6}_{-1.5}$ & - & - & - \\
 & C6 & $10.9^{+0.2}_{-0.2}$ & $6.15^{+0.13}_{-0.14}$ & $175^{+3}_{-3}$ & - & - & - \\
 \enddata
 \tablecomments{Column (1): J2000 name, Column (2): component label matched to Figure \ref{fig:component_fit}, Column (3): intrinsic FWHM major, Column (4): intrinsic FWHM minor, Column (5): low-frequency flux density, Column (6): spectral index, Column (7): component magnetic field strength, Column (8): component minimum energy. CSO-2s without $\alpha$, $B_\mathrm{\mathrm{ME}}$, or $E_{\min}$ are those with single VLBA maps. Components with $\alpha<-1.5$ likely suffer from a significant amount of resolved-out flux, so their magnetic field strengths and energies are likely overestimated and should be taken with caution. Source names that are asterisked are those that \citetalias{2024ApJ...961..242R} suggest are mildly relativistically beamed and should also be regarded with caution. Components corresponding to identified cores do not have a magnetic field strength or energy measurement listed.}
\end{deluxetable*}

\startlongtable
\begin{deluxetable*}{lcccc}
\tablecaption{Spectral Fits
%instantaneous
} \label{tab:spec_results}
\tablehead{
\colhead{Name} & \colhead{$\alpha$} & \colhead{$F_{\nu}(\nu=50\,\mathrm{GHz})$ (mJy)} & \colhead{$B_\mathrm{\mathrm{ME}}$ (mG)} & \colhead{$E_{\min}$ ($\mathrm{M_{\odot}c^2}$)} \\
}
\startdata
J0029+3456 & $-0.50^{+0.02}_{-0.03}$ & $390^{+20}_{-20}$ & $16.3^{+0.2}_{-0.2}$ & $9.5^{+0.3}_{-0.3}$ \\
J0111+3906 & $-1.27^{+0.04}_{-0.04}$ & $103^{+8}_{-8}$ & $97^{+10}_{-8}$ & $11^{+2}_{-2}$ \\
J0405+3803 & $-0.52^{+0.03}_{-0.03}$ & $250^{+20}_{-20}$ & $7.7^{+0.2}_{-0.2}$ & $0.164^{+0.009}_{-0.008}$ \\
J0713+4349 & $-0.86^{+0.05}_{-0.07}$ & $250^{+40}_{-40}$ & $25.4^{+2.3}_{-1.6}$ & $13.7^{+2.3}_{-1.4}$ \\
J0741+2706 & $-0.68^{+0.03}_{-0.03}$ & $122^{+7}_{-6}$ & $8.8^{+0.5}_{-0.4}$ & $35^{+4}_{-3}$ \\
J0855+5751 & $-0.78^{+0.06}_{-0.10}$ & $40^{+8}_{-8}$ & $6.8^{+1.1}_{-0.5}$ & $0.023^{+0.007}_{-0.003}$ \\
J0943+1702 & $-0.46^{+0.04}_{-0.04}$ & $103^{+7}_{-7}$ & $25.9^{+1.1}_{-1.0}$ & $15.4^{+1.1}_{-1.0}$ \\
J1035+5628 & $-1.04^{+0.11}_{-0.11}$ & $130^{+30}_{-20}$ & $29^{+5}_{-4}$ & $14^{+5}_{-3}$ \\
J1111+1955 & $-0.86^{+0.04}_{-0.04}$ & $80^{+6}_{-6}$ & $17.1^{+0.9}_{-0.9}$ & $2.7^{+0.3}_{-0.2}$ \\
J1120+1420 & $-0.92^{+0.02}_{-0.02}$ & $99^{+7}_{-6}$ & $6.7^{+0.3}_{-0.2}$ & $46^{+3}_{-3}$ \\
J1158+2450 & $-1.07^{+0.04}_{-0.04}$ & $161^{+9}_{-9}$ & $29^{+3}_{-2}$ & $3.5^{+0.6}_{-0.5}$ \\
J1159+5820$^{\star}$ & $-0.76^{+0.07}_{-0.07}$ & $67^{+16}_{-13}$ & $13.8^{+1.2}_{-0.9}$ & $47^{+7}_{-5}$ \\
J1227+3635$^{\star}$ & $-1.43^{+0.04}_{-0.04}$ & $19.6^{+1.3}_{-1.4}$ & $45^{+3}_{-3}$ & $750^{+100}_{-90}$ \\
J1244+4048$^{\star}$ & $-0.70^{+0.06}_{-0.06}$ & $130^{+20}_{-20}$ & $7.6^{+0.6}_{-0.5}$ & $66^{+10}_{-7}$ \\
J1247+6723 & $-0.75^{+0.13}_{-0.15}$ & $34^{+12}_{-10}$ & $37^{+7}_{-5}$ & $0.016^{+0.005}_{-0.004}$ \\
J1313+5458 & $-0.91^{+0.06}_{-0.07}$ & $63^{+11}_{-11}$ & $12.0^{+1.1}_{-1.0}$ & $30^{+5}_{-4}$ \\
J1326+3154 & $-0.64^{+0.02}_{-0.03}$ & $510^{+30}_{-40}$ & $8.87^{+0.19}_{-0.17}$ & $22.4^{+0.8}_{-0.7}$ \\
J1400+6210$^{\star}$ & $-0.82^{+0.05}_{-0.05}$ & $250^{+30}_{-30}$ & $7.8^{+0.6}_{-0.5}$ & $60^{+9}_{-7}$ \\
J1407+2827 & $-1.4^{+0.2}_{-0.2}$ & $220^{+70}_{-70}$ & $100^{+30}_{-20}$ & $0.4^{+0.2}_{-0.1}$ \\
J1414+4554 & $-0.63^{+0.06}_{-0.06}$ & $46^{+10}_{-8}$ & $6.7^{+0.3}_{-0.2}$ & $0.69^{+0.05}_{-0.03}$ \\
J1434+4236 & $-1.8^{+0.3}_{-0.5}$ & $6^{+9}_{-4}$ & $26^{+19}_{-9}$ & $300^{+600}_{-200}$ \\
J1440+6108 & $-0.51^{+0.09}_{-0.11}$ & $43^{+14}_{-12}$ & $7.7^{+0.5}_{-0.2}$ & $2.5^{+0.4}_{-0.2}$ \\
J1511+0518 & $-1.10^{+0.05}_{-0.06}$ & $147^{+10}_{-8}$ & $65^{+8}_{-6}$ & $0.11^{+0.03}_{-0.02}$ \\
J1609+2641 & $-1.34^{+0.04}_{-0.05}$ & $64^{+5}_{-5}$ & $25^{+2}_{-2}$ & $114^{+19}_{-14}$ \\
J1644+2536 & $-0.89^{+0.03}_{-0.03}$ & $30^{+3}_{-3}$ & $10.8^{+0.5}_{-0.5}$ & $14.0^{+1.2}_{-1.1}$ \\
J1734+0926 & $-0.88^{+0.04}_{-0.05}$ & $103^{+9}_{-9}$ & $26.6^{+1.9}_{-1.7}$ & $13.8^{+1.9}_{-1.6}$ \\
J1735+5049 & $-0.18^{+0.06}_{-0.07}$ & $590^{+70}_{-70}$ & $100^{+20}_{-20}$ & $12^{+6}_{-4}$ \\
J1816+3457 & $-0.66^{+0.12}_{-0.12}$ & $70^{+30}_{-20}$ & $7.3^{+0.8}_{-0.3}$ & $1.9^{+0.4}_{-0.1}$ \\
J1915+6548 & $-1.2^{+0.3}_{-0.3}$ & $30^{+20}_{-10}$ & $28^{+15}_{-10}$ & $11^{+13}_{-6}$ \\
J1944+5448 & $-0.65^{+0.05}_{-0.05}$ & $180^{+30}_{-30}$ & $11.3^{+0.4}_{-0.3}$ & $2.66^{+0.15}_{-0.10}$ \\
J1945+7055 & $-0.47^{+0.05}_{-0.05}$ & $210^{+30}_{-30}$ & $9.5^{+0.5}_{-0.3}$ & $0.33^{+0.04}_{-0.02}$ \\
J2022+6136 & $-0.56^{+0.05}_{-0.05}$ & $930^{+60}_{-50}$ & $53^{+5}_{-5}$ & $0.66^{+0.12}_{-0.09}$ \\
J2203+1007 & $-1.35^{+0.05}_{-0.05}$ & $21.4^{+1.7}_{-1.6}$ & $45^{+4}_{-4}$ & $72^{+14}_{-12}$ \\
J2355+4950 & $-0.55^{+0.05}_{-0.05}$ & $380^{+50}_{-40}$ & $5.60^{+0.13}_{-0.10}$ & $9.5^{+0.3}_{-0.3}$ \\
\enddata
\tablecomments{Column (1): J2000 name, Column (2): spectral index from radio SED, Column (3): Flux density at 50 GHz, Column (4): minimum energy magnetic field strength, Column (5): minimum energy. We have asterisked the sources that are suggested to be mildly relativistically beamed according to \citetalias{2024ApJ...961..242R}, so their energies should be regarded with caution.}
\end{deluxetable*}

\clearpage
\bibliography{sample7}{}
\bibliographystyle{aasjournalv7}



\end{document}